\documentclass[a4paper,12pt]{article}
\usepackage{comment}   
\usepackage{jheppub}
\usepackage{yfonts}       
\usepackage{amsmath,amssymb,amsthm,amscd,comment,dsfont,graphicx,hyperref,MnSymbol,slashed}
\usepackage{psfrag}
\usepackage{subcaption}
\usepackage[dvipsnames]{xcolor}
\usepackage{soul}
\input epsf.sty
\usepackage{tikz}
\usetikzlibrary{decorations.pathreplacing}

\let\originalleft\left
\let\originalright\right
\renewcommand{\left}{\mathopen{}\mathclose\bgroup\originalleft}
\renewcommand{\right}{\aftergroup\egroup\originalright}

\theoremstyle{definition}

\makeatletter
\newcommand*{\letterdef@}{}
\newcommand*{\letterdef}[3]{%
	\def\letterdef@##1{\expandafter\newcommand\csname #1\endcsname{#2{##1}}}%
	\@tfor\@tempa :=#3\do{\expandafter\letterdef@\expandafter{\@tempa}}}
\makeatother
\letterdef{c#1} {\mathcal}{ABCDEFGHIJKLMNOPQRSTUVWXYZ} 
\letterdef{rm#1}{\mathrm} {dDeimM} 

\def\ie{\begin{equation}\begin{aligned}}
\def\fe{\end{aligned}\end{equation}}

\def\IR{{\mathbb R}}

\newcommand{\tr}{{\rm Tr}}
\newcommand{\re}{{\rm e}}

\newcommand{\ri}{{\rm i}}
\newcommand{\rd}{{\rm d}}

\newcommand{\Tr}{\mathop{\rm Tr}\nolimits}

\newcommand{\bs}{\boldsymbol}

\newcommand{\be}{\begin{equation}}
\newcommand{\ee}{\end{equation}}
\newcommand{\ba}{\begin{aligned}}
\newcommand{\ea}{\end{aligned}}
\newcommand{\ben}{\begin{eqnarray}\displaystyle}
\newcommand{\een}{\end{eqnarray}}

\newcommand{\vev}[1]{{\left\langle #1 \right\rangle}}

\newdimen\tableauside\tableauside=1.0ex
\newdimen\tableaurule\tableaurule=0.4pt
\newdimen\tableaustep
\def\phantomhrule#1{\hbox{\vbox to0pt{\hrule height\tableaurule width#1\vss}}}
\def\phantomvrule#1{\vbox{\hbox to0pt{\vrule width\tableaurule height#1\hss}}}
\def\sqr{\vbox{%
  \phantomhrule\tableaustep
  \hbox{\phantomvrule\tableaustep\kern\tableaustep\phantomvrule\tableaustep}%
  \hbox{\vbox{\phantomhrule\tableauside}\kern-\tableaurule}}}
\def\squares#1{\hbox{\count0=#1\noindent\loop\sqr
  \advance\count0 by-1 \ifnum\count0>0\repeat}}
\def\tableau#1{\vcenter{\offinterlineskip
  \tableaustep=\tableauside\advance\tableaustep by-\tableaurule
  \kern\normallineskip\hbox
    {\kern\normallineskip\vbox
      {\gettableau#1 0 }%
     \kern\normallineskip\kern\tableaurule}%
  \kern\normallineskip\kern\tableaurule}}
\def\gettableau#1{\ifnum#1=0\let\next=\null\else
\squares{#1}\let\next=\gettableau\fi\next}

\definecolor{darkergreen}{rgb}{0.0, 0.6, 0.0}

\preprint{CERN-TH-2025-016 \\ \hspace*{\fill}QMUL-PH-25-03}
\title{\boldmath  Large charge meets semiclassics in $\mathcal{N}=4$ super Yang--Mills}
\author[a]{Augustus Brown,}
\author[a]{Francesco Galvagno,}
\author[b]{Alba Grassi,}
\author[b]{Cristoforo Iossa,}
\author[a]{Congkao Wen}
\affiliation[a]{Centre for Theoretical Physics, Department of Physics and Astronomy, \\
Queen Mary University of London, London, E1 4NS, UK}
\affiliation[b]{Section de Math\'ematiques, Universit\'e de Gen\`eve, 1211 Gen\`eve 4, Switzerland and\\ Theoretical Physics Department, CERN, 1211 Geneva 23, Switzerland }
\affiliation{}

\emailAdd{a.a.x.brown@qmul.ac.uk}
\emailAdd{f.galvagno@qmul.ac.uk}
\emailAdd{Alba.Grassi@unige.ch}
\emailAdd{Cristoforo.Iossa@unige.ch}
\emailAdd{c.wen@qmul.ac.uk}

\abstract{We study the large-charge sector of $\mathcal{N}=4$ super Yang-Mills theory (SYM) with $SU(N)$ gauge group by constructing a special class of half-BPS heavy operators, termed ``canonical operators". Such operators exhibit remarkable simplicity in the large-charge 't Hooft limit, where the dimension of the  operators $\Delta \to \infty$ with $\Delta\, g_{\text{YM}}^2$ held finite. Canonical operator insertions in this regime map $\mathcal{N}=4$ SYM  onto the Coulomb branch, by assigning a classical profile to the scalar fields with non-vanishing values along the diagonals given by the roots of unity. We follow a semiclassical approach to study two-point, three-point and Heavy-Heavy-Light-Light (HHLL) correlators. In particular we show that HHLL correlators in the large-charge 't Hooft limit are computed as two-point functions in a background determined by the classical profiles.  We provide consistent evidence of our findings 
by computing the same observables via supersymmetric localization.}

\begin{document}

\maketitle

\flushbottom
\section{Introduction}
Conformal Field Theories (CFTs) are among the simplest quantum systems, yet they retain a rich structure. A good understanding of these theories thus provides profound insights into the fundamental structure of more general quantum field theories and critical phenomena. A well consolidated strategy to analyze any physical system, and CFTs make no exception, is to make some parameter large. As a parameter grows large, the path integral typically gets dominated by saddle points, and the theory becomes amenable to a semiclassical analysis. A well-known example is the weak coupling expansion, where the saddle point corresponds to the free theory. A more recently understood case is the expansion at large quantum numbers \cite{ HOOFT1974461,Moshe:2003xn,Basso:2006nk, Alday:2007mf, Komargodski:2012ek,Fitzpatrick:2012yx,Berenstein:2002jq}. 
In this limit, massive fluctuations around the saddle become heavy and decouple, reducing the semiclassical description to an Effective Field Theory (EFT) containing  a bunch of massless fields.
Over the past few years, a variety of semiclassical and large charge EFT techniques have been employed to analyze this expansion, making it possible to study operators with large global charges in several classes of CFTs, providing universal predictions for their scaling dimensions and structure constants \cite{Hellerman:2015nra, Monin:2016jmo, Alvarez-Gaume:2016vff, Badel:2019oxl,Cuomo:2022kio,Cuomo:2021ygt}.

In this context a special role is played by superconformal field theories (SCFTs) with a space of vacua, whose large charge EFT captures low-energy physics on the Coulomb branch. Such correspondence has been widely tested in the context of 4d $\mathcal{N}=2$ SCFTs at rank-1, by comparing EFT predictions with the results for extremal correlators coming from supersymmetric localization \cite{Hellerman:2017sur,Hellerman:2018xpi,Grassi:2019txd,Bourget:2018obm,Beccaria:2018xxl,Beccaria:2020azj}\footnote{See also \cite{Heckman:2024erd,Baume:2022cot,Baume:2020ure} for investigations of the large charge expansion in the context of the 6d $\mathcal{N}=(2,0)$ and $\mathcal{N}=(1,0)$ theory.}. A first step towards the generalization to higher rank in  $\mathcal{N}=2$ SCFTs was undertaken in \cite{Grassi:2024bwl} although a semiclassical  interpretation of these results is still lacking.

The connection between large charge insertions and Coulomb branch physics has since then moved to the simplest gauge theory in 4d, i.e. $\cN=4$ Supersymmetric Yang-Mills (SYM). The rank-1 case was analysed in \cite{Caetano:2023zwe}. A class of half-BPS Heavy-Heavy-Light-Light (HHLL) correlators was studied in the so-called large-charge 't Hooft limit \cite{Bourget:2018obm}, proving that it can be interpreted as a two-point function of the light operators (chosen as the superconformal primaries in the stress tensor multiplet) in the classical background generated by the large-charge insertions. 
In this limit the coupling gets smaller as the charge grows larger keeping their product fixed, and  massive fluctuations around the large charge saddle point do not decouple unless the large charge 't Hooft coupling is sent to infinity.
A closed form for this particular HHLL correlator was obtained to all orders in perturbation theory, also leading to a completely resummed expression. 

Going beyond rank-1 theories is a challenging task due to the large degeneracy of heavy operators.
A first attempt at studying large charge in $\cN=4$ SYM was made within the framework of integrated correlators. These correlators are constructed by integrating out the spacetime dependence using a specific supersymmetry-preserving measure \cite{Binder:2019jwn}, and can be calculated by supersymmetric localization \cite{Moore:1997dj,Lossev:1997bz,Pestun:2007rz}.
 For a special choice of the heavy operators, namely the maximal trace operators and their relatives,\footnote{The maximal trace operators are the superconformal primaries that, at a fixed conformal dimension, have the maximal number of traces. In $SU(N)$ gauge theories they correspond to $\left( T_2(x,Y) \right)^p$, following the notation of \eqref{eq:Op_def}, and the large-charge limit implies taking $p \rightarrow \infty$. A similar behaviour in this particular large-charge limit is assumed by operators of the form $\left( T_2(x,Y) \right)^p  \mathcal{O}$, where one takes $p \rightarrow \infty$ while the dimension of the additional component $\mathcal{O}$ is kept finite.} 
the integrated correlators can be determined recursively using the so-called Laplace-difference equations \cite{Paul:2022piq, Brown:2023cpz}. The large-charge expansion, even at finite Yang-Mills coupling, has been systematically developed using various techniques \cite{Paul:2023rka, Brown:2023why}. This framework has uncovered several remarkable properties, such as identifying a specific class of non-holomorphic modular functions that govern the large-charge expansion for integrated correlators of this class of heavy operators.

The HHLL correlators with insertions of generic heavy half-BPS operators have been studied in \cite{Brown:2024yvt} perturbatively in the large-charge 't Hooft limit. 
Building on the results of \cite{Beccaria:2020azj} and utilizing the chiral Lagrangian insertion mechanism~\cite{Eden:2012tu}, it was shown that the HHLL correlators take a universal form which completely fixes the spacetime dependence to all orders, up to certain color factor coefficients that depend on the precise form of the heavy operators (see \autoref{sec:largecharge} for a brief review). 
It was suggested in \cite{Brown:2024yvt} that there exist special classes of heavy operators (called ``canonical operators'' in that reference), for which the color factor coefficients significantly simplify and allow for an all-order resummation.  Such canonical behavior was explicitly demonstrated for gauge groups $SU(2)$ and $SU(3)$. The case of the $SU(2)$ gauge group agrees with the results of \cite{Caetano:2023zwe} obtained by a saddle point analysis. 
However, a systematic study of the `canonical operators' for higher rank $SU(N)$ remained elusive.

A complementary approach in the understanding of large charge integrated correlators was developed in \cite{Grassi:2024bwl}, where the $SU(3)$ case was analyzed in detail. Building on \cite{Grassi:2019txd}, it was demonstrated that, in the large charge regime, the integrated correlators are effectively described by a combination of Wishart and Jacobi matrix models, coupled in a non-trivial manner. An interesting outcome of this analysis was the identification of a special subset of operators in higher-rank theories whose behavior closely mimics that of half-BPS operators in rank-1 theories: these are precisely the `canonical operators' appearing in \cite{Brown:2024yvt}. 

The aim of this paper is to generalize and systematize the study of canonical operators for a generic $SU(N)$ gauge group, going beyond the case of integrated correlators and providing a semiclassical computational framework. An explicit semiclassical analysis in the spirit of \cite{Hellerman:2015nra, Monin:2016jmo, Alvarez-Gaume:2016vff, Badel:2019oxl,Caetano:2023zwe} is challenging due to the intricate structure of the UV Lagrangian, the complexity of canonical operators, and the rich dynamics of the Coulomb branch physics. Therefore, in this paper, we tackle the problem by adopting the following strategy\footnote{Of course, in real life the order was the opposite: starting from \cite{Brown:2024yvt,Grassi:2024bwl}, we identified the canonical operators, computed their correlation functions via localization and diagrammatic analysis, and then reconstructed their semiclassical description.}:
\begin{itemize}
\item[-] we conjecture a semiclassical profile for the six scalars in $\mathcal{N}=4$ SYM;
\item[-] studying fluctuations around such a configuration, we determine the mass spectrum on the Coulomb branch;
\item[-] we compute the correlators from a path integral approach;
\item[-] we extensively check our results with a large charge diagrammatic analysis and via supersymmetric localization.
\end{itemize}
We now provide a detailed outline of the paper, also summarizing the main results.


\subsection*{Outline and main results}

In \autoref{sec:definition} we introduce the notations for operators and correlators in $\mathcal{N}=4$ SYM and define canonical operators, denoted by $\mathcal{O}_K^m(x,Y)$ for $3 \leq K \leq N $. They belong to an orthogonal basis of the chiral ring of half-BPS superconformal primaries; their R-charge and scaling dimension $\Delta$ are related as
\be 
\label{eq:dimD}\Delta = m K= {R\over 2}~,\ee
where $m$ and $K$ identify their trace numbers and trace structure respectively (see around equation \eqref{eq:OmK} for the explicit definition).
In this paper we investigate correlation functions involving two canonical operators  $\mathcal{O}_K^m(x_i, Y_i)$ with  $i=1,2$  in the large-charge double-scaling limit: 
\be\label{eq:thin}
m  \rightarrow \infty \, ,  \qquad \text{with} \qquad \lambda = {m \,  g^2_{_{\rm YM}} \over 16 \pi^2 }
\quad \text{fixed} \, . 
\ee
We refer to this scaling as a large charge ’t Hooft limit, where the rank of the gauge group remains fixed. 
The limit \eqref{eq:thin} is different from the pure large charge limit (where the charge of the operator is taken to infinity while keeping the other parameters fixed, including $g_{_{\rm YM}}$), as it retains the non-trivial coupling dependence of the observables already at the first nontrivial order in $m$.

In \autoref{sec:semi-ana} we study $\mathcal{N}=4$ SYM in presence of canonical operators via a semiclassical analysis.
As explained in \autoref{sec:saddle_mass}, when canonical operators become heavy they source the following semiclassical background for the six scalars:
\begin{equation} \label{eq:Zclass0}
     \langle \Phi_I (x) \rangle_{\theta} = \frac{ 2\sqrt{\lambda} }{ \sqrt{d_{12}}} \left( e^{i\theta} \Omega_K^{(N)} \frac{(Y_1)_I}{(x-x_1)^2} + e^{-i\theta} \overline{\Omega_K^{(N)}} \frac{(Y_2)_I}{(x-x_2)^2} \right)  \,, \quad ~I=1,\dots, 6\,. 
\end{equation}
Here $ d_{ij}= Y_i\cdot Y_j/x_{ij}^2$, $\theta\in [0,2\pi]$ is the modulus of the solutions that we need to integrate over \cite{Monin:2016jmo}, and $\Omega_K^{(N)}$ is an $N\times N$ diagonal matrix of the form
\be\label{eq:Zclass-intr}
\Omega^{(N)}_K = 
\begin{pmatrix}  
\omega^{(K)}_1 & 0 & \cdots & 0 & \cdots & \,\, 0 \\
0 & \omega^{(K)}_2 & \cdots & 0 & \cdots & \,\, 0 \\
\vdots & \vdots & \ddots & \vdots & \cdots & \, \,\vdots \\
0 & 0 & \cdots & \omega^{(K)}_K & \cdots & \,\, 0 \\
\vdots & \vdots & \vdots & \vdots & \ddots & \,\, 0 \\
0 & 0 & \cdots & 0 & \cdots & \,\, 0
\end{pmatrix} \, , \qquad {\rm with} \quad \omega_n^{(K)} = e^{\frac{2\pi i}{K} n} \, .
\ee
The vev \eqref{eq:Zclass0} is responsible for the following gauge symmetry breaking pattern
\be\label{eq:symbint}
SU(N) \longrightarrow U(1)^{K-1} \times U(N-K)~.
\ee
Accordingly, fluctuations around \eqref{eq:Zclass0} are described by $(K-1)+(N-K)^2$ massless scalars corresponding to unbroken directions, while the remaining $2K(N-K)+K(K-1)$ scalars acquire a mass $M_s$ due to the Higgs mechanism. 

In \autoref{sec:2pteft} and \autoref{sec:semi_3pt} we compute half-BPS two- and three- point functions assuming the semiclassical profile \eqref{eq:Zclass0}.
These correlators are protected by supersymmetry \cite{Lee:1998bxa,Bianchi:1999ie, Baggio:2012rr} and are tree-level exact.
However, the dependence on $N$ and the charge $2mK$ is non-trivial. Let us note that, as coupling independent observables, in these cases the large charge 't Hooft limit is equivalent to a pure large charge limit. We find that the two-point functions of canonical operators $\mathcal{O}_K^m$ in the large-charge limit take the form\footnote{We adopt a normalization for the chiral primaries where the OPE is simple, but the two-point functions are non-trivial. This is the usual normalization used in the context of $\cN=2$ extremal correlators, see e.g. \cite[eq.~(3.2)]{Gerchkovitz:2016gxx} or \cite[pg.~3]{Grassi:2019txd}.}: 
\begin{equation}\label{eq:tpoN_intreo}
    \langle \mathcal{O}^m_K(x_1,Y_1)\mathcal{O}^m_K(x_2,Y_2) \rangle = \left( \frac{d_{12}}{4\pi^2} \right)^{\Delta} \Gamma(\Delta+1) \left( \Delta\right)^{\alpha_{N,K}} A^{\Delta} B \left(1+ {O}\left({1\over m}\right)\right)\,,
\end{equation}
where  $\Delta$ is the dimension of $\mathcal{O}^m_K$ as given in \eqref{eq:dimD}. The parameters $A$ and $B$ are unphysical normalization factors, while $\alpha_{N,K}$ is a physical parameter that 
accounts for the number of massless degrees of freedom around \eqref{eq:Zclass0}. We find
\be \label{eq:alphaintro}\alpha_{N,K}= {K\over 2}(2N-K-1)~.\ee
The result \eqref{eq:tpoN_intreo} closely resembles the EFT prediction for extremal correlators in rank-1 SCFT obtained in \cite{Hellerman:2017sur}. Moreover, in the maximally symmetry breaking case $K=N$, we obtain an expression to all orders in $1/m$ which closely resembles the one in \cite{Hellerman:2018xpi}. We find 
\begin{equation}\label{eq:tpoN_intreo1}
    \langle \mathcal{O}^m_N(x_1,Y_1) \, \mathcal{O}^m_N(x_2,Y_2) \rangle = \left( \frac{d_{12}}{4\pi^2} \right)^{\Delta} f_{N}(m) 2^{-\Delta} N^{2m-\Delta} \frac{\Gamma\left(\Delta +\alpha_{N,N}+1\right)}{\Gamma\left(\alpha_{N,N}+1\right)}\,,
\end{equation}
where $ f_{N}(m)= a(N)+\mathcal{O}(\re^{-m})$. Examples of $ f_{N}(m)$ are given in \eqref{eq:fexa}. Let us nevertheless emphasize that, a priori, it is far from obvious why the same EFT of \cite{Hellerman:2017sur,Hellerman:2018xpi} should still hold in higher-rank cases.

In the large charge limit, where canonical operators become heavy, we find that three-point functions can be interpreted as one-point functions evaluated in the background induced by the heavy operators
\begin{equation}\label{eq:intro3p-intr}
    \frac{\vev{\mathcal{O}^m_K(x_1, Y_1) \mathcal{O}^m_K(x_2, Y_2) \cO(x_3, Y_3)}}{\vev{\mathcal{O}^m_K(x_1, Y_1) \mathcal{O}^m_K(x_2, Y_2)}}
\simeq  \int_0^{2\pi} {d\theta \over 2\pi}\langle \cO(x_3,Y_3) \rangle_{{\theta}} \,,
\end{equation}
where  $\langle \dots \rangle_{\theta}$ denotes the correlator on the Coulomb Branch in the background \eqref{eq:Zclass0}. Moreover $\simeq$ means equality at leading order in the scaling limit \eqref{eq:thin}. We will use this $\simeq$ notation throughout the paper.

In \autoref{sec:4pt_free}, \autoref{sec:4pt_interact} and \autoref{sec:4pt_ope} we present another key result of this paper: the computation of four-point correlators involving two canonical operators and two light operators of dimension two. This is an observable of great interest that receives quantum corrections at all orders, making a direct computation highly challenging.\footnote{For example, the perturbative expansion of related correlators in the standard large-$N$ planar limit have been computed explicitly only to three loops \cite{Drummond:2013nda}, even though the loop integrands have been determined to higher-loop orders \cite{Bourjaily:2016evz, Caron-Huot:2021usw, He:2024cej}.} When considering the large-charge double-scaling
limit \eqref{eq:thin}, this observable is once again completely determined by the classical profile \eqref{eq:Zclass0} to any orders in the coupling $\lambda$. In particular, the dynamical part of the correlator $\cT_{m,K}$ containing all the coupling dependence, as defined in  \eqref{eq:4ptHH22}, takes the following form: 
\be\label{eq:4pts}
\mathcal{T}_{m,K}\left(u,v; \lambda \right) \simeq \frac{1}{2 u} \sum_s \left( \left( \sum_{\ell= 0}^{\infty} (-  M_s^2 )^{\ell}\,  P^{(\ell)}(u, v)\right)^2 - 1\right)~,
\ee
where $P^{(\ell)}(u, v)$ is the $\ell-$loop ladder Feynman integral,  which is known to any loop order \cite{Usyukina:1993ch}, as given in \eqref{eq:ladderL}. This combination corresponds to a product of two scalar massive propagators of mass $M_s$, and can also be found in the context of large charge limit in $O(N)$ field theory \cite{Giombi:2020enj} and in the rank-1 case in $\cN=4$ SYM \cite{Caetano:2023zwe}. 
The sum over $s$ runs over the massive spectrum corresponding to the symmetry breaking pattern \eqref{eq:symbint}.
The masses $M_s$  can be read off explicitly from the vev \eqref{eq:Zclass0}.  There are \( K(N-K) \) of them, all with mass
    \begin{equation}
        M = \sqrt{\lambda}\, , 
    \end{equation}
and additional ones with mass  \begin{equation}\label{eq:aijmass-intr}
        M^{(ij)}  = 2 \sqrt{\lambda}\,  \sin\left(\frac{i {-} j}{K}\pi\right)\, , \qquad 0 \leq i < j < K \, .
    \end{equation}
By substituting the expression of the masses in \eqref{eq:4pts}, we obtain the explicit form of four-point correlators. 
Remarkably, the ladder Feynman integrals can be resummed  \cite{Broadhurst:2010ds} as in \eqref{eq:resum-form}, and so the expression of the four-point correlators \eqref{eq:4pts} is valid even non-perturbatively in $\lambda$.

In \autoref{sec:localization} we  move to the sphere and introduce matrix model and supersymmetric localization techniques, and define the $S^4$ equivalent of canonical operators. In \autoref{sec:testloc} we use such matrix model techniques from localization to extensively check the full set of results derived from semiclassical arguments in the previous sections. We also provide some explicit results for the four-point functions beyond perturbation theory and we comment on the combination of large-charge with large-$N$ limits. Finally, \autoref{sec:conc} contains our conclusions and the perspectives for future work, and all the technical material about localization computations is detailed in three appendices.

\section{Canonical operators and their correlation functions}\label{sec:definition}

We study classes of half-BPS superconformal primary operators built from single trace combinations of the $\cN=4$ scalar fields $\Phi^I$ with $I=1,\dots,6,$ 
\begin{equation}\label{eq:Op_def}
 T_p(x, Y)  =  \left(\frac{\tau_2}{4\pi}\right)^{p/2} \, Y_{I_1} \cdots Y_{I_p} \Tr\left( \Phi^{I_1}(x) \cdots \Phi^{I_p}(x) \right)~, 
\end{equation}
where $\tau_2$ is the purely imaginary part of the complexified Yang-Mills coupling: 
\begin{equation}\label{eq:tau_complex}
    \tau:=\tau_1+ \ri \tau_2 = \frac{\theta}{2\pi} +\ri \frac{4\pi}{g_{_{\rm YM}}^2}~.
\end{equation} 
In equation \eqref{eq:Op_def}, $x$ is the spacetime position of the operator and $Y_I$ is a null vector, which carries the $SO(6)$ R-symmetry indices. Throughout this paper we consider correlation functions of heavy operators, whose conformal dimension $\Delta$ is the largest parameter of the theory: in particular, $\Delta \gg N^2$.
Due to the $SU(N)$ trace relations, such operators are necessarily made of combinations of multi-trace operators, which form a basis of local half-BPS superconformal primary operators. Multi-trace operators are defined as 
\begin{equation}
    \label{eq:mul-tr}
 T_{{\bf m}}(x, Y) =  \prod_{k\geq 1} T_{m_k}(x, Y)  \, , 
\end{equation}
where 
\be {\bf m} = \{m_1, m_2 \cdots\}\, , \quad {\rm with} \quad m_i\geq m_{i+1} ~,\quad m_i\in \{2,\dots, N\} ~.\ee
Their scaling dimension $\Delta_{\bf m}$ and $R$-charge $R_{\bf m}$ are related by  
\be  R_{\bf m}=2\Delta_{\bf m}= 2 \sum_{k\geq 1} \, m_k ~.
\ee
At a given scaling dimension $\Delta$ there is a large degeneracy of multitrace operators that are not orthogonal with respect to each other, that is 
\be\label{eq:mat2ptT}
\langle T_{\bf{m}}(x_1, Y_1) T_{\bf{m}'}(x_2, Y_2) \rangle = \mathcal{N}_{\bf{m} \bf{m}'} (N,K)\left(\frac{d_{12}}{4\pi^2} \right)^{\Delta_{\bf{m}}} \,, \quad \sum_{k=1} m_k = \sum_{k=1} m_k'
\ee
where we introduced the notation $d_{ij}$ for the free propagator
\be
d_{ij} = \frac{Y_i\cdot Y_j}{x_{ij}^2}\, , \qquad {\rm with} \qquad x_{ij}:= x_i-x_j \, . 
\ee
The coefficient $\mathcal{N}_{\bf{m} \bf{m}'}$ is a non-diagonal matrix whose entries can be computed from localization, as we will see in the following. Note that, since two- and three-point functions of half-BPS operators are protected, $\mathcal{N}_{\bf{m} \bf{m}'}$ does not receive any quantum corrections. It is then convenient to construct a diagonal basis obtained through a Gram-Schmidt procedure on the matrix $\mathcal{N}_{\bf{m} \bf{m}'}$. To this end, we order operators of a given dimension $\Delta$ as follows: operators with the fewest traces of $2$ in $T_{\bf{m}}$ come first, and if these are the same, then the operators with the fewest traces of $3$ in $T_{\bf{m}}$ come first, and so on up to traces of $N$.
 For example at $\Delta = 8$, $N=5$ we have 
\begin{equation}\label{eq:op105}
    \{ T_{4,4}(x,Y) \,, \,\, T_{5,3}(x,Y) \,, \,\, T_{3,3,2}(x,Y) \,, \,\, T_{4,2,2}(x,Y) \,, \,\, T_{2,2,2,2}(x,Y) \} \,.
\end{equation}
After a Gram-Schmidt diagonalization on \eqref{eq:mat2ptT}, we obtain 
\allowdisplaybreaks{
\begin{align}\label{eq:ex55}
    \mathcal{O}_{4,4}(x,Y) &= T_{4,4}(x,Y)+ \sum_{{\bf{n}} \ne(4,4)} c_{\bf{n}}^{(4,4)} T_{\bf{n}}(x,Y) \,, \cr
    \mathcal{O}_{5,3}(x,Y) &= T_{5,3}(x,Y)+ \sum_{{\bf{n}} \ne(4,4),(5,3)} c_{\bf{n}}^{(5,3)} T_{\bf{n}}(x,Y)  \,, \\
    \vdots   \nonumber
\end{align}
where the coefficients $c_{\bf n}$ of the linear combinations ensure that
\begin{equation}\label{eq:Omk_orthogonal}
    \langle \mathcal{O}_{\bf{m}}(x_1,Y_1) \, \mathcal{O}_{\bf{m}'}(x_2,Y_2) \rangle \propto \delta_{\bf{m} \bf{m}'} \,.
\end{equation}
The coefficients $c_{\bf n}$ only depend on $N$ and can be more conveniently computed by using the matrix model on the four-sphere, as detailed in \autoref{sec:localization}. In particular we refer to \autoref{sec:canonical} for some explicit examples.

Among these operators $\mathcal{O}_{\bf{m}}$, the {\it canonical operators} in particular will play a crucial role in the following sections. Canonical operators are defined as\footnote{It is worth noting that when $m=1$ the operators constructed here are essentially the so-called single-particle operators \cite{Aprile:2020uxk}. However, in this paper we are interested in the large-charge limit with $m \to \infty$.} 
\be\label{eq:OmK}  
\mathcal{O}_{\!\!\! \begin{tikzpicture}[baseline=(O.base)]
    \node (O) {\footnotesize{$K, \dots, K$}};
    \draw [decorate,decoration={brace,mirror,amplitude=5pt,raise=0.5pt}]
        (O.south west) -- (O.south east) node[midway,below=3pt] {\footnotesize $m$};
\end{tikzpicture}}\!\! (x,Y) := \mathcal{O}^m_K(x,Y) \,, \qquad K>2 \,.
\ee
For future reference, their two-point function takes the form  
\begin{align} \label{eq:2pt_pp}
    \vev{ \mathcal{O}^m_K (x_1, Y_1) \mathcal{O}^m_K(x_2, Y_2)} =\cN_{m}(N,K) \left(\frac{d_{12}}{4\pi^2} \right)^{mK} \,,
\end{align}
where $\cN_{m}(N,K)$ is a normalization constant.

In the following sections we will study the three- and a particular set of four- point functions involving two canonical operators with large dimension and $R-$charge. Let us start by discussing the general structure of these correlators. The normalized three-point function of two canonical operators and one multitrace operator $T_{\bf p}$ takes the form:
\begin{align}\label{eq:3pt_HHp}
\frac{\vev{\mathcal{O}^m_K(x_1,Y_1) \mathcal{O}^m_K(x_2,Y_2) T_{\bf p}(x_3,Y_3)} }{\vev{\mathcal{O}^m_K(x_1,Y_1) \mathcal{O}^m_K(x_2,Y_2)}}= \mathfrak{C}_{m m {\bf p}}(N,K)\times \left( \frac{1}{2\pi^2} \frac{d_{23}d_{31}}{d_{12}}\right)^{\frac{\Delta_{\bf p}}{2}}\,,
 \end{align}
where  $\mathfrak{C}_{mm{\bf p}}$ are the normalized three-point coefficients, and the last term corresponds to the spacetime and R-symmetry kinematic factor, which is fixed by superconformal symmetry. 

As for the four-point function, we consider correlation functions of two canonical operators and two single-trace operators of the form \eqref{eq:Op_def} with dimension $p=2$.
Normalizing by the two point function of canonical operators, any four-point function of this kind can be expressed as 
\begin{equation}\label{eq:4ptHH22}
     \frac{\vev{\mathcal{O}^m_K(x_1, Y_1) \mathcal{O}^m_K(x_2, Y_2) T_2(x_3, Y_3) T_2(x_4, Y_4) }}{\vev{\mathcal{O}^m_K(x_1, Y_1) \mathcal{O}^m_K(x_2, Y_2)}} = \cG_{\mathrm {free}}(x_i,Y_i) + \cI_4(x_i,Y_i) \,  \cT_{m,K}(u,v; \tau, \bar{\tau}) ~,
\end{equation}
where $\cG_{\mathrm {free}}$ is the free theory correlator that can be computed via Wick contraction and the rest of the RHS contains all the quantum corrections. The factor $\cI_4(x_i,Y_i)$ contains the full R-symmetry dependence of the quantum corrections and is completely fixed by superconformal symmetry \cite{Eden:2000bk, Nirschl:2004pa} as follows:
\begin{equation}\label{eq:I4_Rsym}
    \cI_4(x_i,Y_i) =\left(\frac{d_{34}}{4 \pi^2}\right)^{2} \frac{(z-\alpha)(z-\bar \alpha)(\bar z-\alpha)(\bar z-\bar \alpha)\, z \bar{z}}{\alpha^2  \,\bar{\alpha}^2\, (1-z)(1-\bar{z})} \, .
\end{equation}
The conformal invariant cross-ratios are defined as follows
\ie \label{eq:zzb}
   u &=z\, \bar{z}= {x_{12}^2 x_{34}^2 \over x_{13}^2 x_{24}^2} ~, \quad \quad v=(1-z)(1- \bar{z})= {x_{14}^2 x_{23}^2 \over x_{13}^2 x_{24}^2} ~,
   \fe 
whereas the R-symmetry cross-ratios are given by
   \ie \label{eq:Rsymm_crossrat}
   \alpha\, \bar{\alpha}&=  {Y_1\cdot Y_2 Y_3\cdot Y_4 \over Y_1\cdot Y_3 Y_2\cdot Y_4} ~, \quad \quad (1-\alpha)(1- \bar{\alpha})=  {Y_1\cdot Y_4 Y_2\cdot Y_3 \over Y_1\cdot Y_3 Y_2\cdot Y_4}  ~.
\fe
We mainly concentrate on the dynamical part of the four-point functions, which after the factorization of \eqref{eq:4ptHH22} is fully contained in the so-called reduced correlator $\cT_{m, K}(u,v)$.  More precisely, we are interested in the correlators in the large charge double scaling limit
\begin{equation}\label{eq:thooft20}
    m \to \infty \,,\quad \tau_2 \to \infty \,, \quad\lambda = {g_{\rm YM}^2 m \over 16 \pi^2} = \frac{ m}{4\pi \tau_2} \quad \text{fixed}~.
\end{equation}
We refer to this limit as the large-charge 't Hooft limit.

\section{A semiclassical analysis}
\label{sec:semi-ana}

In this section we compute correlation functions involving two canonical operators $\mathcal{O}^m_K$ in the large charge 't Hooft limit  \eqref{eq:thooft20} from semiclassical considerations.  
Here we outline the analysis of the saddle point equation for the path integral and present the expression for the semiclassical solution, which provides the basic ingredient for the computation of
correlation functions in the large charge 't Hooft limit. The complete derivation of the semiclassical background and EFT for canonical operators as well as other classes of operators for $SU(N)$ gauge group is beyond the scope of this paper. In the upcoming sections, \autoref{sec:localization} and \autoref{sec:testloc}, we will provide non-trivial checks of the saddle solution by computing the correlators through other means.

\subsection{Saddle-point solution and mass spectrum}\label{sec:saddle_mass}

We consider the correlation functions of two canonical operators and some other operators with fixed conformal dimensions, 
\be\label{eq:effsadd}
\vev{\mathcal{O}_K^m(x_1, Y_1)\mathcal{O}_K^m(x_2, Y_2) \ldots } = \frac{1}{\mathcal{Z}} \int \mathcal{D} [\text{fields}] \, \re^{-S_{\text{SYM}}}\mathcal{O}_K^m(x_1, Y_1)\mathcal{O}_K^m(x_2,  Y_2 ) \ldots ~ .
\ee
In the large charge 't Hooft limit, the path integral is dominated by the saddle points, which are the solution to a saddle point equation of the type:
\ie \label{eq:effsadd2}
\delta_{\text{fields}} \left( -S_{\text{SYM}} +  \log(\mathcal{O}_K^m(x_1, Y_1)\mathcal{O}_K^m(x_2,  Y_2) )  \right) =0 \, .
\fe
It is straightforward to observe that a solution to the above saddle-point equation should behave as $\Phi^{\rm cl}\sim\sqrt{\lambda}$, where $\lambda$ is defined in \eqref{eq:thooft20}. This scaling suggests that the large-charge 't Hooft limit is a well-motivated and appropriate regime to consider. However, finding the explicit solution is highly non-trivial, especially  due  to the complicated construction of $\mathcal{O}_K^m$ (with $m\to \infty$) as described in the previous section. As we will test extensively in the forthcoming sections, we find that inserting two heavy canonical operators in the path integral of $\mathcal{N}=4$ SYM sources a nontrivial profile for the six scalars $\Phi_I$ that can be computed from a saddle point approximation. More precisely, we claim that the saddle point equation of \eqref{eq:effsadd2} in the limit \eqref{eq:thooft20} is solved by
\begin{equation} \label{eq:Zclass2}
     \Phi^{\text{cl}}_I (x)  = \frac{2\sqrt{\lambda}}{\sqrt{d_{12}}} \left( e^{i\theta} \Omega_K^{(N)}  {(Y_1)_I \over (x-x_1)^2} + e^{-i\theta} \overline{\Omega_K^{(N)}}  {(Y_2)_I  \over (x-x_2)^2 }\right) \, , 
\end{equation}
where $\theta \in [0,2\pi]$ is the modulus of the solutions, and $\Omega_K^{(N)}$ is an $N\times N$ diagonal matrix of the form
\be\label{eq:Zclass}
\Omega^{(N)}_K = 
\begin{pmatrix}  
\omega^{(K)}_1 & 0 & \cdots & 0 & \cdots & \,\, 0 \\
0 & \omega^{(K)}_2 & \cdots & 0 & \cdots & \,\, 0 \\
\vdots & \vdots & \ddots & \vdots & \cdots & \, \,\vdots \\
0 & 0 & \cdots & \omega^{(K)}_K & \cdots & \,\, 0 \\
\vdots & \vdots & \vdots & \vdots & \ddots & \,\, 0 \\
0 & 0 & \cdots & 0 & \cdots & \,\, 0
\end{pmatrix}, \qquad {\rm with} \quad \omega_n^{(K)} = e^{\frac{2\pi i}{K} n} \,.
\ee
The spacetime dependence of $ \Phi^{\text{cl}}_I (x) $ is fixed by dimensional analysis and conformal invariance. 
This vev for the scalars effectively brings the theory to a Coulomb phase corresponding to the symmetry breaking pattern
\be
SU(N) \longrightarrow U(1)^{K-1} \times U(N-K) \,, \quad 2 < K \le N \,.
\ee
Fluctuations around this saddle are described by an EFT containing 
\be\label{eq:masseft}
\underbrace{(K-1)}_{U(1)^{K-1}} + \underbrace{(N-K)^2}_{U(N-K)}
\ee
massless $\mathcal{N}=4$ multiplets corresponding to the unbroken directions. To better characterize fluctuations around the background \eqref{eq:Zclass}, it is convenient to introduce the $\mathcal{N}=4$ complex scalars 
\be
Z =  \frac{\Phi_1 + i \Phi_2}{\sqrt{2}} \,, \quad Y = \frac{\Phi_3+i\Phi_4}{\sqrt{2}} \,, \quad X = \frac{\Phi_5+i\Phi_6}{\sqrt{2}} \,~.
\ee
Without loss of generality we can take $Y_1=\overline{Y_2}=\{1,-i,0,0,0,0\}$ in \eqref{eq:Zclass2}.
With this choice, the saddle point equation at large charge give a vanishing profile to $X$ and $Y$ and nontrivial profile to $Z$
\be
X^{\rm cl} (x)= 0~,~~~~Y^{\rm cl} (x)=0~,~~~~ Z^{\rm cl} (x) = 2 \sqrt{\lambda}\, e^{i\theta} \Omega_K^{(N)} \frac{|x_1-x_2|}{(x-x_1)^2}  \, , 
\ee 
and the conjugate fields are given by
\be
\overline{X}^{\rm cl} (x)= 0~,~~~~ \overline{Y}^{\rm cl} (x)=0~,~~~~ \overline{Z}^{\rm cl} (x) = 2 \sqrt{\lambda}\, e^{-i\theta} \overline\Omega_K^{(N)} \frac{|x_1-x_2|}{(x-x_2)^2}  \,.
\ee
Note that $Z^{\rm cl}$ and $\overline{Z}^{\rm cl}$ are not complex conjugate to each other on this nontrivial saddle.

We now expand the full $\mathcal{N}=4$ action around the classical vev as\footnote{The $\mathcal{N}=4$ SYM action has to be completed with ghosts and a gauge fixing terms. We refer to \cite{Caetano:2023zwe,Ivanovskiy:2024vel} for conventions.}
\begin{equation}
    \Phi_I = \Phi_I^{\rm cl} + \delta\Phi_I~,
\end{equation}
and concentrating in particular on the scalar quartic term $\tfrac{1}{2}[\Phi_I,\Phi_J]^2$, we get the following mass terms for scalar fluctuations 
\be
S \supset \frac{2}{g_{_{\rm YM}}^2} \int d^4 x \frac{1}{4} \Tr \left( |[\delta X, Z^{\rm cl}]|^2 + |[\delta Y, Z^{\rm cl}]|^2 + |[\delta Z , Z^{\rm cl}]|^2\right) \,.
\ee
As anticipated, modes along the unbroken direction remain massless. All off-diagonal modes out of the $(N-K)\times(N-K)$ block corresponding to the unbroken $U(N - K)$ get a spacetime dependent mass given by
\be \label{eq:mij}
m_{ij}(x)^2 = |Z^{\rm cl}_i- Z^{\rm cl}_j|^2 = {4(x_1 -x_2)^2 \over (x-x_1)^2 (x-x_2)^2} M_{ij}^2 \,, \quad M_{ij} = \sqrt{\lambda} \,  \left|\omega_i^{(K)}- \omega_j^{(K)} \right| \,. 
\ee
Note that the masses do not depend on the modulus $\theta$. They can be classified as follows:
\begin{enumerate} 
    \item There are $2K(N-K)$ excitations, all with mass
    \begin{equation}\label{eq:aijmass0}
        M_{ij} = \sqrt{\lambda} \,, \quad i=1, \dots, K \,, \,\, j=K+1, \dots, N \,.
    \end{equation}
    \item There are an additional $K(K-1)$ excitations with masses
    \begin{equation}\label{eq:aijmass}
        M_{ij} = 2 \sqrt{\lambda} \sin\left(\frac{i-j}{K}\pi\right), \quad 1 \leq i < j \leq K \, .
    \end{equation}
 \end{enumerate}
As a sanity check, let us note that
\be
\underbrace{(K-1)+(N-K)^2}_{\rm massless} \, +\,  \underbrace{2K(N-K)}_{M_{ij} = \sqrt{\lambda}}\, +\!\!\!  \underbrace{K(K-1)}_{M_{ij}=2\sqrt{\lambda}  \sin\left(\frac{s}{K}\pi\right)} \!\!\! = N^2-1 \,.
\ee
We remark that the masses given in \eqref{eq:aijmass} can be reorganized in slightly nicer form, and  we conclude the structure of the masses is given by 
\ie \label{eq:Ms}
    M_s  &=  2 \sqrt{\lambda} \, \sin\left(\frac{s}{K}\pi\right), \quad \, {\rm with} \quad 1 \le s < K \,, \qquad {\rm appear} \,\,\,\, K \,\,\,\, {\rm times} \, , \cr 
    M_s  &= \sqrt{\lambda}  \,, \qquad \qquad \quad ~ ~\,  {\rm with}\quad s=K \, , \quad\quad\quad \,\,\, {\rm appear} \,\,\,\,\, 2K(N{-}K) \,\,\,\, {\rm times} \, .
\fe
The first line in the above equation provides a rewritten form of \eqref{eq:aijmass}, while the second line includes the terms from \eqref{eq:aijmass0}. 
These masses for various values of $K$ are displayed in \autoref{fig:mass2}. For further reference, we note that $M_s$'s satisfy the following simple properties
\be\label{eq:summssq}
\begin{aligned}
&\sum_s M_s^2 = 2 K N \lambda \,, \\
&\sum_s  \log (M_s) = K\left(\log K+\left(N-\frac{K+1}{2}\right)\log \lambda \right) \,,
\end{aligned}
\ee
where all the masses have been counted with their multiplicity.  

\begin{figure}[t]
\centering
\begin{subfigure}[b]{0.4\textwidth}
\begin{center}
    \begin{tikzpicture}
        \draw[->] (-0.2,0) -- (3.5,0) node[right] {$s$};
        \draw[->] (0,-0.2) -- (0,3.5) node[right] {$M_s/\sqrt{\lambda}$};

        \draw[thick] (3,0) arc[start angle=0, end angle=90, radius=3cm];
        \draw[dashed, orange] (0,0) -- (45:3cm);
        \draw[dashed, lightgray] (0,{3*0.707}) -- ({3*0.707},{3*0.707});
        \fill[orange] (0,{3*0.707}) circle (1.5pt);
        \draw[dashed, red] (0,0) -- (30:3cm);

        \draw[dashed, lightgray] (0,{3*0.5}) -- ({3*0.866},{3*0.5});

        \fill[red] (0,{3*0.5}) circle (1.5pt);

        \draw[dashed, red] (0,0) -- (60:3cm);

        \draw[dashed, lightgray] (0,{3*0.866}) -- ({3*0.5},{3*0.866});

        \fill[red] (0,{3*0.866}) circle (1.5pt);

        \draw[dashed, blue] (0,0) -- (18:3cm);
        \draw[dashed, lightgray] (0,{3*0.31}) -- ({3*0.95},{3*0.31});
        \fill[blue] (0,{3*0.31}) circle (1.5pt);
        
        \draw[dashed, blue] (0,0) -- (36:3cm);
        \draw[dashed, lightgray] (0,{3*0.59}) -- ({3*0.81},{3*0.59});
        \fill[blue] (0,{3*0.59}) circle (1.5pt);
        
        \draw[dashed, blue] (0,0) -- (54:3cm);
        \draw[dashed, lightgray] (0,{3*0.81}) -- ({3*0.59},{3*0.81});
        \fill[blue] (0,{3*0.81}) circle (1.5pt);
        
        \draw[dashed, blue] (0,0) -- (72:3cm);
        \draw[dashed, lightgray] (0,{3*0.95}) -- ({3*0.31},{3*0.95});
        \fill[blue] (0,{3*0.95}) circle (1.5pt);

        \fill[black] (0,3) circle (1.5pt);

    \end{tikzpicture}
    \caption{$K=$ \textcolor{orange}{4}, \textcolor{red}{6}, \textcolor{blue}{8}}
    \end{center}
\end{subfigure}
\begin{subfigure}[b]{0.4\textwidth}
\begin{center}
    \begin{tikzpicture}
        \draw[->] (-0.2,0) -- (3.5,0) node[right] {$s$};
        \draw[->] (0,-0.2) -- (0,3.5) node[right] {$M_s/\sqrt{\lambda}$};

        \draw[thick] (3,0) arc[start angle=0, end angle=90, radius=3cm];
        
        \draw[dashed, violet] (0,0) -- (8.57:3cm);
        \draw[dashed, lightgray] (0,{3*0.15}) -- ({3*0.99},{3*0.15});
        \fill[violet] (0,{3*0.15}) circle (1.5pt);
        
        \draw[dashed, violet] (0,0) -- (17.14:3cm);
        \draw[dashed, lightgray] (0,{3*0.295}) -- ({3*0.95},{3*0.295});
        \fill[violet] (0,{3*0.295}) circle (1.5pt);
        
        \draw[dashed, violet] (0,0) -- (25.71:3cm);
        \draw[dashed, lightgray] (0,{3*0.434}) -- ({3*0.90},{3*0.434});
        \fill[violet] (0,{3*0.434}) circle (1.5pt);
        
        \draw[dashed, violet] (0,0) -- (34.29:3cm);
        \draw[dashed, lightgray] (0,{3*0.563}) -- ({3*0.83},{3*0.563});
        \fill[violet] (0,{3*0.563}) circle (1.5pt);

        \draw[dashed, violet] (0,0) -- (42.86:3cm);
        \draw[dashed, lightgray] (0,{3*0.680}) -- ({3*0.733},{3*0.680});
        \fill[violet] (0,{3*0.680}) circle (1.5pt);

        \draw[dashed, violet] (0,0) -- (51.43:3cm);
        \draw[dashed, lightgray] (0,{3*0.782}) -- ({3*0.623},{3*0.782});
        \fill[violet] (0,{3*0.782}) circle (1.5pt);

        \draw[dashed, violet] (0,0) -- (60:3cm);
        \draw[dashed, lightgray] (0,{3*0.866}) -- ({3*0.5},{3*0.866});
        \fill[violet] (0,{3*0.866}) circle (1.5pt);

        \draw[dashed, violet] (0,0) -- (68.57:3cm);
        \draw[dashed, lightgray] (0,{3*0.931}) -- ({3*0.365},{3*0.931});
        \fill[violet] (0,{3*0.931}) circle (1.5pt);

        \draw[dashed, violet] (0,0) -- (77.14:3cm);
        \draw[dashed, lightgray] (0,{3*0.975}) -- ({3*0.223},{3*0.975});
        \fill[violet] (0,{3*0.975}) circle (1.5pt);

        \draw[dashed, violet] (0,0) -- (85.71:3cm);
        \draw[dashed, lightgray] (0,{3*0.997}) -- ({3*0.0747},{3*0.997});
        \fill[violet] (0,{3*0.997}) circle (1.5pt);

    \end{tikzpicture}
       \caption{$K$= \textcolor{violet}{21}}
       \end{center}
\end{subfigure}
\caption{Values of the $K(K-1)$ masses described in the first line of \eqref{eq:Ms} (or equivalently \eqref{eq:aijmass})  for various values of $K$. As $K$ becomes larger, the interval $[0,2]$ on the $M_s/\sqrt{\lambda}$ axis becomes densely covered, for example $K=21$ as shown in figure (b).}
\label{fig:mass2}
\end{figure}
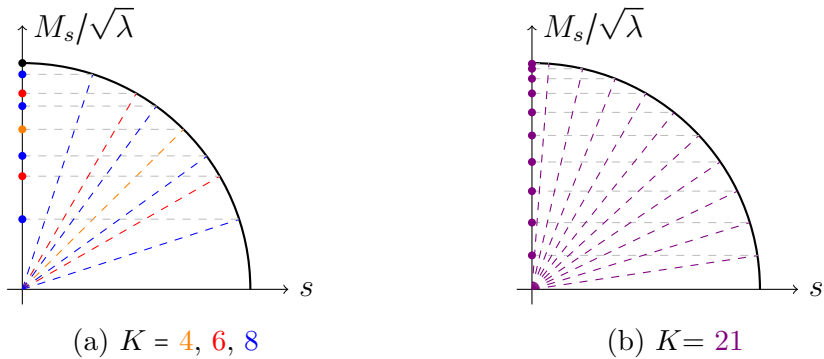

We now move to the computation of correlators starting from the semiclassical profile \eqref{eq:Zclass2}. We will then extensively verify our findings using localization techniques in \autoref{sec:localization}.

\subsection{Two-point functions}\label{sec:2pteft}
We begin by examining the two-point functions. As two-point correlators of canonical operators are extremal,
one might anticipate that they exhibit a structure analogous to that found in rank-1 SCFTs \cite{Hellerman:2017sur}. Recent results for rank-2 \cite{Grassi:2024bwl} support this expectation, here we extend that analysis to general rank. 
In the large charge limit\footnote{As mentioned in the introduction, since these two-point functions are tree-level exact, there is no difference between the double scaling large charge limit \eqref{eq:thooft20}, and the pure large charge limit in which $\tau_2$ is kept fixed, discussed in e.g. \cite{Hellerman:2017sur}.} we then expect two point functions \eqref{eq:2pt_pp} to take the following universal form
\be 
\label{eq:log2pn}\log \cN_{m}(N,K) =\log\left(\Gamma(\Delta {+} 1)\right)+\alpha_{N,K} \log (\Delta)+ \Delta \log( A)+B+ {O}\left( \Delta^{-1}\right)\, , 
\ee
where $\Delta=mK$ is the dimension of the canonical operator $\cO_{m}^K$, and $A,B$ are unphysical normalization factors.  The factor $\alpha_{N,K}$ on the other hand is physical, and is given by
\be 
\alpha_{N,K} = 2\Delta a=2(a_{\rm CFT}-a_{\rm EFT}) \, ,  \ee
where $a_{\rm CFT}$ is the $a$-anomaly coefficient of the full interacting CFT and $a_{\rm EFT}$ is the $a$-anomaly coefficient of the EFT describing fluctuations around \eqref{eq:Zclass}\cite{Hellerman:2017sur}. 
Following \cite{Anselmi:1997ys} (see also \cite[Appendix A]{Hellerman:2017sur}), every $\mathcal{N}=4$ massless multiplet contributes to the $a$-anomaly with a factor of $1/4$. In the CFT we have $N^2-1$ massless multiplets corresponding to the unbroken $SU(N)$ generators, therefore
\be a_{\rm CFT}={1\over 4}(N^2-1). \ee
On the other hand, from \eqref{eq:masseft} we find
\be a_{\rm EFT}= {1\over 4} \left(K-1+(N-K)^2\right) \,.\ee
All in all this gives
\be\label{eq:alpha} \alpha_{N,K}={K\over 2}\left(2N-K-1\right) \,. \ee
Let us remark that this can be contrasted with two-point functions of maximal trace operators \cite{Gerchkovitz:2016gxx} 
\be
\vev{T_2^m(x_1,Y_1) T_2^m(x_2,Y_2)} = \left(\frac{d_{12}}{4\pi^2}\right)^{2m} 2^{2m} \Gamma(m+1) \frac{\Gamma\left(\frac{N^2-1}{2}+m\right)}{\Gamma\left(\frac{N^2-1}{2}\right)}\,.
\ee
Taking the large charge limit, we can read off $\alpha$ to be\footnote{We thank Luigi Tizzano for bringing this to our attention.}
\be
\alpha_{T_2^m} = \frac{N^2-2}{2} = 2\left(\frac{N^2-1}{4}-\frac{1}{4} \right) \,.
\ee
This suggests that fluctuations around the saddle point selected by heavy maximal trace operators would be described by an EFT containing one massless $\mathcal{N}=4$ multiplet. 

\subsection{Three-point functions}\label{sec:semi_3pt}

We now turn our attention to three-point functions of two heavy canonical and one light operator. Again this observables are tree-level exact, so the large-charge double scaling limit \eqref{eq:thooft20} coincides with the pure large charge limit. The light operator does not enter the saddle point equation, therefore our prediction for the scalar three-point function in the large-charge double scaling limit  is
\begin{equation}\label{eq:intro3p}
    \frac{\vev{\mathcal{O}^m_K(x_1, Y_1) \mathcal{O}^m_K(x_2, Y_2) \mathcal{O}(x_3,Y_3)}}{\vev{\mathcal{O}^m_K(x_1, Y_1) \mathcal{O}^m_K(x_2, Y_2)}}
\simeq \int_0^{2\pi} \frac{d\theta}{2\pi} \langle \mathcal{O}(x_3,Y_3) \rangle_{{\theta}} \,,
\end{equation}
where  $\langle \dots \rangle_{\theta}$ is the correlator on the Coulomb Branch where the scalar acquires the vev given by the classical solution \eqref{eq:Zclass}
\begin{equation} 
    \langle \Phi_I (x) \rangle_{\theta} = \Phi_I^{\rm cl} (x) \,,
\end{equation}
and we integrate over the saddle point modulus $\theta$. The semiclassical computation \eqref{eq:intro3p} reproduces the expected shape of the three-point function \eqref{eq:3pt_HHp} and provides the leading order in the large charge limit of the three-point coefficient $\mathfrak{C}_{mm \bs p}$.

Let us start with the simplest example by considering $\mathcal{O}(x_3)\equiv T_{2} (x_3, Y_3)$. Following \eqref{eq:intro3p} at leading order in $m^{-1}$ expansion, we have: 
\begin{equation}\label{eq:T2example}
     \vev{T_{2} (x_3, Y_3)}_{{\theta}} = \frac{\tau_2}{4\pi} Y_{3}^{I}Y_{3}^{J} \tr \left(  \Phi_I^{\rm cl} (x_3) \Phi_J^{\rm cl} (x_3)\right) = \frac{1}{4\pi^2} \frac{m}{d_{12}} \tr \left( d_{13} \Omega_K^{(N)} e^{i\theta} + d_{23} \overline{\Omega_K^{(N)}} e^{-i\theta}\right)^2~.
\end{equation}
The integration over $\theta$ sets to zero all the terms containing oscillating $e^{ik\theta}$ factors for any integer $k$. The only $\theta$-independent term reconstructs the R-symmetry/spacetime structure of the three-point function \eqref{eq:3pt_HHp}. Moreover since $\tr (\Omega_K^{(N)} \overline{\Omega_K^{(N)}}) = K$ we get:
\begin{equation}
    \mathfrak{C}_{mm 2} \simeq  mK~.
\end{equation}
The generalization to the case $\mathcal{O}(x_3)\equiv T_{\bf p} (x_3, Y_3)$ is straightforward. We have 
\be
\langle T_{\bf p} (x_3, Y_3) \rangle_{{\theta}}  =\langle \prod_{k\geq 1} T_{p_k} (x_3, Y_3) \rangle_{{\theta}} =\prod_{k\geq 1} \langle T_{p_k} (x_3, Y_3) \rangle_{{\theta}} \, ,
\ee
where
\be  \langle T_{p} (x_3, Y_3) \rangle_{{\theta}}= \left(\frac{\tau_2}{4\pi}\right)^{p/2}  Y_{3}^{I_1}\dots Y_{3}^{I_p} \text{Tr} \left( \langle \Phi_{I_1} (x_3) \rangle_{\theta} \dots \langle \Phi_{I_p} (x_3) \rangle_{\theta} \right) \, . \ee
Again the integration over $\theta$ kills oscillating terms, therefore we have 
\be\label{eq:zeroint}
\int_0^{2\pi} \frac{d\theta}{2\pi} \langle T_{\bf p} (x_3, Y_3) \rangle_{{\theta}} = \left(\frac{1}{2\pi^2} \frac{d_{23} d_{31}}{d_{12}}\right)^{\frac{\Delta_{\bf p}}{2}}  \left(\frac{m}{2}\right)^{\frac{\Delta_{\bf p}}{2}} \int_0^{2\pi} \frac{d \theta}{2\pi} \prod_k \Tr \left(e^{i\theta} \Omega^{(N)}_K+ \text{c.c.}\right)^{p_k} \,,
\ee
since the difference in the integrands is set to zero by the $\theta-$integral. Furthermore, since 
\be\label{eq:trace_vanishing}
\Tr \left(\Omega_K^{(N)}\right)^\ell = 0\,,\quad \ell= 1, \dots, K-1 \, ,
\ee
we have, for $p_k \leq K$,
\be
\Tr \left(e^{i\theta} \Omega^{(N)}_K+ \text{c.c.}\right)^{p_k} = \begin{cases}
 K \left(e^{i K \theta} + e^{-i K \theta} \right)\delta_{p_k K} \,, \qquad\qquad\quad \text{$K$ odd} \,, \\ K \binom{p_k}{p_k/2} + K \left(e^{i K \theta} + e^{-i K \theta} \right)\delta_{p_k K} \,, \quad \text{$K$ even} \,.
\end{cases}
\ee
Note that the integral will vanish if there are any odd traces in $\textbf{p}$ for $p_k < K$. For odd $K$, with $K$ appearing $n$ times in $\textbf{p}$ and all remaining traces even (and less than $K$), we simply get 
\be \label{eq:threeptodd}
\mathfrak{C}_{m m {\bf p}}(N,K) \simeq \left(\frac{m}{2}\right)^{\frac{\Delta_{\bf p}}{2}} 
K^n \binom{n}{n/2} \frac{1+(-1)^n}{2} \left(\prod_{\text{even} \, p_k } K \binom{p_k}{\frac{p_k}{2}} \right) \,.
\ee
For even $K$, the integral is slightly more complicated. Let $n$ be again the number of traces in $\textbf{p}$ where $p_k =K$. Then
\begin{equation}
    K^n\left( \binom{K}{K/2} + \left(e^{i K \theta} + e^{-i K \theta} \right) \right)^n = K^n \sum_{b=0}^n \binom{n}{b} \binom{K}{K/2}^{n-b} \left(e^{iK\theta}+e^{-iK\theta}\right)^b \,.
\end{equation}
The integral will set to zero all $\theta-$dependent terms, therefore 
\be
\int_0^{2\pi} \frac{d \theta}{2\pi} \sum_{b=0}^n \binom{n}{b} \binom{K}{K/2}^{n-b} \left(e^{iK\theta}+e^{-iK\theta}\right)^b = \sum_{b'=0}^{n/2} \binom{n}{2b'} \binom{K}{K/2}^{n-2b'} \binom{2b'}{b'} \,.
\ee
The last sum is an hypergeometric sum. All in all we get 
\be
\int_0^{2\pi} \frac{d \theta}{2\pi} \prod_{k=1}^n \Tr \left(e^{i\theta} \Omega^{(N)}_K+ \text{c.c.}\right)^{K} = K^n \binom{K}{K/2}^n {}_2 F_1 \left(\frac{1-n}{2},-\frac{n}{2},1,\frac{4}{\binom{K}{K/2}^2}\right) \,,
\ee
and more generally (where again, all remaining traces are even and less than $K$), we get for even $K$ 
\be\label{eq:threepteven}
\mathfrak{C}_{m m {\bf p}}(N,K) \simeq \left(\frac{m}{2}\right)^{\frac{\Delta_{\bf p}}{2}}   \left(\prod_{\text{even} \, p_k < K} K \binom{p_k}{\frac{p_k}{2}} \right) K^n \binom{K}{K/2}^n {}_2 F_1 \left(\frac{1-n}{2},-\frac{n}{2},1,\frac{4}{\binom{K}{K/2}^2}\right) \,.
\ee
Note that single trace factors with $p_k>K$ are related to multi trace operators with $p_k\le K$ by $SU(K)$ trace relations, since the upper $K\times K$ block in \eqref{eq:Zclass} is an $SU(K)$ matrix.

\subsection{HHLL four-point functions: free-theory part}\label{sec:4pt_free}

The profile configuration given in \eqref{eq:Zclass} allows us to compute the four-point HHLL correlators by interpreting it as a two-point function in a heavy background. The action of superconformal symmetry constrains the half-BPS four-point function as displayed in \eqref{eq:4ptHH22}.
We start by computing the leading contribution in the large charge limit $m\to\infty$ to the free-theory part of the correlator $\cG_{\mathrm {free}}(x_i,Y_i)$ following the semiclassical analysis. The general free theory part of the four-point function can be determined by Wick contractions and is decomposed into the following six R-symmetry channels: 
\begin{align}\label{eq:4pt_free}
    \cG_{\mathrm {free}}= \left(\frac{d_{34}}{4\pi^2}\right)^2\left(\frac{F_1}{\alpha^2\bar\alpha^2} +  \frac{(1{-}\alpha)(1{-}\bar\alpha)}{\alpha^2\bar\alpha^2} F_2 + \frac{(1{-}\alpha)^2(1{-}\bar\alpha)^2}{\alpha^2\bar\alpha^2} F_3 + \frac{F_4}{\alpha\bar\alpha}  + \frac{(1{-}\alpha)(1{-}\bar\alpha)}{\alpha\bar\alpha} F_5 + F_6 \right) \,,
\end{align}
where $\alpha,\, \bar\alpha$ are the R-symmetry cross-ratios defined in \eqref{eq:Rsymm_crossrat} and the functions $F_i(u,v)$ are functions of the spacetime cross-ratios \eqref{eq:zzb}. The six channels are displayed in \autoref{fig:free_channels}. For our four-point function in the presence of canonical operators, it turns out that the channels $F_1$ and $F_3$ vanish identically. Indeed, both $F_1$ and $F_3$ are proportional to $\langle \cO_K^m T_2 \cO_{mK - 2} \rangle$, where $\cO_{mK - 2}$ is some operator with dimension $mK-2$. Then the orthogonality condition \eqref{eq:Omk_orthogonal} implies  the vanishing of $F_1$ and $F_3$.\footnote{An analogous phenomenon happens to four-point functions $\langle \cO_s \cO_s T_2 T_2 \rangle$, where $\cO_s$ are the so-called single particle operators~\cite{Aprile:2020uxk}, which obey similar orthogonality properties as \eqref{eq:Omk_orthogonal}. See for example the appendix A of \cite{Alday:2019nin}.} As we will see, the semiclassical computation will confirm this cancellation at least at leading order in large charge.

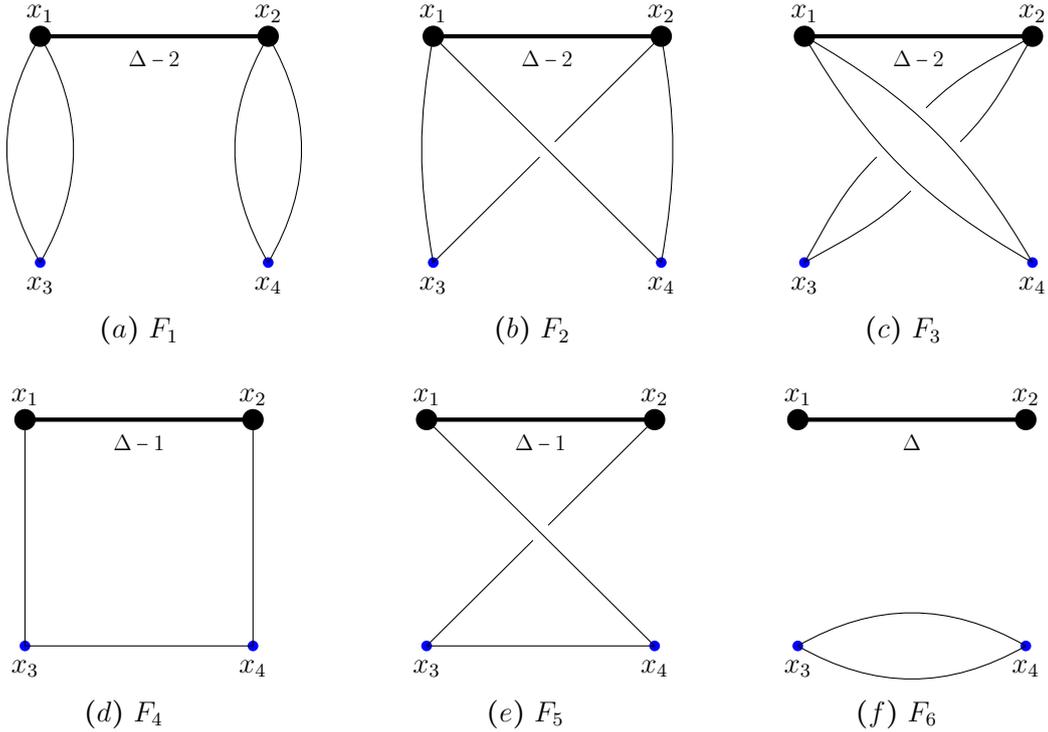
\begin{figure}[t!]
    \centering
\begin{tikzpicture}
    \coordinate (x1) at (-1.5,1.5);
    \coordinate (x2) at (1.5, 1.5);
    \coordinate (x3) at (-1.5, -1.5);
    \coordinate (x4) at (1.5, -1.5);
    
    \fill[blue] (x3) circle (2pt);
    \fill[blue] (x4) circle (2pt);
    \fill[black] (x1) circle (4pt);
    \fill[black] (x2) circle (4pt);
    
    \draw[black][ultra thick] (x1) -- (x2);
    \draw[black] (x1) to[out=-60, in=60] (x3);
    \draw[black] (x1) to[out=-120, in=120] (x3);
    \draw[black] (x2) to[out=-60, in=60] (x4);
    \draw[black] (x2) to[out=-120, in=120] (x4);

    \node[scale=0.9] at (-1.5,1.8) {$x_1$}; \node[scale=0.9]  at (1.5,1.8) {$x_2$};
    \node[scale=0.7] at (0,1.2) {$\Delta-2$};
      \node[scale=0.9]  at (-1.5,-1.8) {$x_3$}; \node[scale=0.9]  at (1.5,-1.8) {$x_4$};
      \node[scale=0.9] at (-0.2,-2.4) {$(a)~ F_1$};
\end{tikzpicture}
\hspace{1cm}
\begin{tikzpicture}
    \coordinate (x1) at (-1.5,1.5);
    \coordinate (x2) at (1.5, 1.5);
    \coordinate (x3) at (-1.5, -1.5);
    \coordinate (x4) at (1.5, -1.5);
    
    \fill[blue] (x3) circle (2pt);
    \fill[blue] (x4) circle (2pt);
    \fill[black] (x1) circle (4pt);
    \fill[black] (x2) circle (4pt);
    
    \draw[black][ultra thick] (x1) -- (x2);
    \draw[black] (x1) to (x4);
    \draw[black] (x1) to[out=-100, in=100] (x3);
    \draw[black] (x2) to[out=-80, in=80] (x4);
    \draw[black] (x2) to (0.1,0.1); \draw[black] (-0.1,-0.1) to (x3);

    \node[scale=0.9] at (-1.5,1.8) {$x_1$}; \node[scale=0.9]  at (1.5,1.8) {$x_2$};
    \node[scale=0.7] at (0,1.2) {$\Delta-2$};
      \node[scale=0.9]  at (-1.5,-1.8) {$x_3$}; \node[scale=0.9]  at (1.5,-1.8) {$x_4$};
      \node[scale=0.9] at (-0.2,-2.4) {$(b)~ F_2$};
\end{tikzpicture}
\hspace{1cm}
\begin{tikzpicture}
    \coordinate (x1) at (-1.5,1.5);
    \coordinate (x2) at (1.5, 1.5);
    \coordinate (x3) at (-1.5, -1.5);
    \coordinate (x4) at (1.5, -1.5);
    
    \fill[blue] (x3) circle (2pt);
    \fill[blue] (x4) circle (2pt);
    \fill[black] (x1) circle (4pt);
    \fill[black] (x2) circle (4pt);
    
    \draw[black][ultra thick] (x1) -- (x2);
    \draw[black] (x1) to[out=-30, in=120] (x4);
    \draw[black] (x1) to[out=-60, in=150] (x4);
    \draw[black] (x2) to[out=-150, in=45] (0.1,0.55);
    \draw[black] (x2) to[out=-120, in=45] (0.55,0.1);
    \draw[black] (-0.55,-0.1) to[out=-135, in=60] (x3);
    \draw[black] (-0.1,-0.55) to[out=-135, in=30] (x3);

    \node[scale=0.9] at (-1.5,1.8) {$x_1$}; \node[scale=0.9]  at (1.5,1.8) {$x_2$};
    \node[scale=0.7] at (0,1.2) {$\Delta-2$};
      \node[scale=0.9]  at (-1.5,-1.8) {$x_3$}; \node[scale=0.9]  at (1.5,-1.8) {$x_4$};
      \node[scale=0.9] at (-0.2,-2.4) {$(c)~ F_3$};
\end{tikzpicture} \\[0.3cm]

\begin{tikzpicture}
    \coordinate (x1) at (-1.5,1.5);
    \coordinate (x2) at (1.5, 1.5);
    \coordinate (x3) at (-1.5, -1.5);
    \coordinate (x4) at (1.5, -1.5);
    
    \fill[blue] (x3) circle (2pt);
    \fill[blue] (x4) circle (2pt);
    \fill[black] (x1) circle (4pt);
    \fill[black] (x2) circle (4pt);
    
    \draw[black][ultra thick] (x1) -- (x2);
    \draw[black] (x1) to (x3);
    \draw[black] (x2) to (x4);
    \draw[black] (x3) to (x4);

    \node[scale=0.9] at (-1.5,1.8) {$x_1$}; \node[scale=0.9]  at (1.5,1.8) {$x_2$};
    \node[scale=0.7] at (0,1.2) {$\Delta-1$};
      \node[scale=0.9]  at (-1.5,-1.8) {$x_3$}; \node[scale=0.9]  at (1.5,-1.8) {$x_4$};
      \node[scale=0.9] at (-0.2,-2.4) {$(d)~ F_4$};
\end{tikzpicture}
\hspace{1.4cm}
\begin{tikzpicture}
    \coordinate (x1) at (-1.5,1.5);
    \coordinate (x2) at (1.5, 1.5);
    \coordinate (x3) at (-1.5, -1.5);
    \coordinate (x4) at (1.5, -1.5);
    
    \fill[blue] (x3) circle (2pt);
    \fill[blue] (x4) circle (2pt);
    \fill[black] (x1) circle (4pt);
    \fill[black] (x2) circle (4pt);
    
    \draw[black][ultra thick] (x1) -- (x2);
    \draw[black] (x1) to (x4);
    \draw[black] (x3) to (x4);
    \draw[black] (x2) to (0.1,0.1); \draw[black] (-0.1,-0.1) to (x3);

    \node[scale=0.9] at (-1.5,1.8) {$x_1$}; \node[scale=0.9]  at (1.5,1.8) {$x_2$};
    \node[scale=0.7] at (0,1.2) {$\Delta-1$};
      \node[scale=0.9]  at (-1.5,-1.8) {$x_3$}; \node[scale=0.9]  at (1.5,-1.8) {$x_4$};
      \node[scale=0.9] at (-0.2,-2.4) {$(e)~ F_5$};
\end{tikzpicture}
\hspace{1cm}
\begin{tikzpicture}
    \coordinate (x1) at (-1.5,1.5);
    \coordinate (x2) at (1.5, 1.5);
    \coordinate (x3) at (-1.5, -1.5);
    \coordinate (x4) at (1.5, -1.5);
    
    \fill[blue] (x3) circle (2pt);
    \fill[blue] (x4) circle (2pt);
    \fill[black] (x1) circle (4pt);
    \fill[black] (x2) circle (4pt);
    
    \draw[black][ultra thick] (x1) -- (x2);
    \draw[black] (x3) to[out=30, in=150] (x4);
    \draw[black] (x3) to[out=-30, in=-150] (x4);

    \node[scale=0.9] at (-1.5,1.8) {$x_1$}; \node[scale=0.9]  at (1.5,1.8) {$x_2$};
    \node[scale=0.7] at (0,1.2) {$\Delta$};
      \node[scale=0.9]  at (-1.5,-1.8) {$x_3$}; \node[scale=0.9]  at (1.5,-1.8) {$x_4$};
      \node[scale=0.9] at (-0.2,-2.4) {$(f)~ F_6$};
\end{tikzpicture}
\caption{The six R-symmetry channels from the free theory Wick contractions. Each thin black line represents a free propagator, the thick line between $x_1$ and $x_2$ stands for the combination of Wick contractions of a number of strands indicated by the number below it. Different contractions between the heavy and the light operators determine different scaling in the $m\to\infty$ limit. In particular, $F_{1,2,3}$ channels scale like $\binom{\Delta}{2} \sim  \tfrac{m^2}{2}$, $F_{4,5}$ channels scale like $m$, the fully disconnected term $F_6$ behaves as $m^0$.} 
    \label{fig:free_channels}
\end{figure}

We can explicitly derive the six channels at leading order in the large charge limit from different semiclassical computations. We recall the expression for the vev for a scalar in the semiclassical background created by the canonical operators with $m\to\infty$ inserted in $x_1$ and $x_2$:
\begin{equation} 
    \langle \Phi_I (x) \rangle_{\theta} = \frac{ 2 \sqrt{\lambda} }{ \sqrt{d_{12}}} \left( e^{i\theta} \Omega_K^{(N)} \frac{(Y_1)_I}{(x-x_1)^2} + e^{-i\theta} \overline\Omega_K^{(N)} \frac{(Y_2)_I}{(x-x_2)^2} \right)  \,,
\end{equation}
In particular the $O(m^2)$ of the correlator comes from the factorized three-point functions
\begin{equation}
   \cG_{\mathrm {free}}|_{O(m^2)} = \int_0^{2\pi} \frac{d\theta}{2\pi} \vev{T_2(x_3,Y_3)}_{\text{cl}} \vev{T_2(x_4,Y_4)}_{\text{cl}} \, , 
\end{equation}
and using the result for $\langle \Phi_I (x) \rangle_{\theta}$ we perform the contractions. After using eq. \eqref{eq:T2example} and integrating over $\theta$ we find
\begin{equation}
    \begin{split}
        \label{eq:F123_semi}
   \int_0^{2\pi} \frac{d\theta}{2\pi} \vev{T_2}_{\text{cl}} \vev{T_2}_{\text{cl}}  =&\,   \left(\frac{1}{4\pi^2d_{12}} \right)^2 m^2  \bigg[(d_{13}^2d_{24}^2+d_{14}^2d_{23}^2)  \tr[(\Omega_K^{(N)})^2] \tr[(\overline\Omega_K^{(N)})^2] \\  &+ 4 d_{13} d_{24} d_{14} d_{23} \left[\tr(\Omega_K^{(N)}\overline\Omega_K^{(N)})\right]^2 \bigg]~.
    \end{split}
\end{equation}
 The first line of \eqref{eq:F123_semi} vanishes due to the trace properties in \eqref{eq:trace_vanishing}. Hence semiclassics imposes 
\begin{equation}
    F_1(u,v)=F_3(u,v)=0\, ,
\end{equation}as expected also from field theory considerations. This is an interesting consistency check of the validity of the semiclassical computation. By comparing \eqref{eq:F123_semi} with \eqref{eq:4pt_free}, we can read:
\begin{equation}
    F_2(u,v) \simeq  4 m^2 K^2   \frac{u^2}{v}~,
\end{equation}
at the leading order in large charge.

The result for $F_{4,5}$ arises from the partial contraction of the $T_2$'s with a free propagator:
\begin{equation}
   \cG_{\mathrm {free}}|_{O(m)} ={ \tau_2 \over 4\pi }\, \frac{d_{34}}{4\pi^2} (Y_3)_{I} (Y_4)_{J} \int_0^{2\pi} \frac{d\theta}{2\pi} \tr \left( \vev{\Phi^{I}(x_3)}_{\text{cl}} \vev{\Phi^{J}(x_4)}_{\text{cl}} \right)~.
\end{equation}
Again after the $\theta$-integration we get:
\begin{equation}
    \cG_{\mathrm {free}}|_{O(m)} = \frac{m}{(4\pi^2)^2}\,\frac{d_{34}}{d_{12}} (d_{13} d_{24} + d_{14} d_{23} ) \tr(\Omega_K^{(N)}\overline\Omega_K^{(N)})~,
\end{equation}
from which, comparing with \eqref{eq:4pt_free} we read:
\begin{equation}
    F_4(u,v) \simeq  m K \, u ~, \qquad F_5(u,v) \simeq  m K \, \frac{u }{v}~,
\end{equation}
again at leading order in large charge.
Finally, $F_6$ comes from the completely factorized form and it is simply a constant for any values of the charge:
\begin{equation}\label{eq:F611}
    F_6(u, v) =  \frac{N^2-1}{2} \, . 
\end{equation}
Collecting all the terms, we get 
\begin{align}\label{eq:4pt_freesubs}
    \cG_{\mathrm {free}} \simeq  \left(\frac{d_{34}}{4\pi^2}\right)^2\left(\frac{(1-\alpha)(1-\bar\alpha)}{\alpha^2\bar\alpha^2} \frac{4 \Delta^2 \, u^2}{v}  + \frac{ \Delta \, u}{\alpha\bar\alpha}  + \frac{(1-\alpha)(1-\bar\alpha)}{\alpha\bar\alpha}  \, \frac{ \Delta \, u}{v} + \frac{N^2-1}{2} \right) \,,
\end{align}
where we recall that $\Delta = mK$. We conclude that the semiclassical analysis represents a valuable tool to compute the leading-$m$ contribution for each R-symmetry channel. Let us emphasize again that here we have only computed the leading-$m$ contribution for each R-symmetry channel.

\subsection{HHLL four-point functions: interacting terms}\label{sec:4pt_interact}

Let us now compute the reduced correlator $\mathcal{T}_{m,K}(u,v;\tau,\bar{\tau})$ which contains all the quantum corrections to the four-point function, as defined in \eqref{eq:4ptHH22}. Thanks to the fact that the reduced correlator is independent of $SO(6)$ R-symmetry for the correlator we consider, we can make a convenient choice of the $SO(6)$ null vectors without loss of generality. In particular, we take  
\be
Y_1=\overline{Y_2}=Y=\left(1,-i,0,0,0,0\right), \qquad Y_3=\overline{Y_4}=Y_X=\left(0,0,0,0,1,i\right) \,.
\ee
This choice R-symmetry factors sets 
\be\label{eq:factXZ}
\mathcal{G}_{\rm free} = \left(\frac{1}{4\pi^2}\right)^2 \frac{4}{x_{34}^4} F_6 \,, \qquad \mathcal{I}_4(x_i, Y_i) = \left(\frac{1}{4\pi^2}\right)^2 \frac{4 u}{x_{34}^4} \,,
\ee
therefore the full correlator \eqref{eq:4ptHH22} reduces to
\be
\frac{\vev{\mathcal{O}^m_K(x_1, Y) \mathcal{O}^m_K(x_2, \overline{Y}) T_2(x_3, Y_X) T_2(x_4, \overline{Y_X})}}{\vev{\mathcal{O}^m_K(x_1, Y) \mathcal{O}^m_K(x_2, \overline{Y})}} = \left(\frac{1}{4\pi^2}\right)^2 \frac{4}{x_{34}^4} \left(F_6 +u \mathcal{T}_{m,K}(u,v;\tau,\bar{\tau}) \right) \,.
\ee
At leading order in the large charge 't Hooft limit, this is given by
\be
\int_0^{2\pi} \frac{d\theta}{2\pi} \langle \Tr X^2(x_3) \Tr \overline{X}^2(x_4) \rangle_{{\theta}} \, . 
\ee
Note that $\langle \Tr X^2(x_3) \rangle_\theta=0$, therefore only connected contractions will contribute to the integral. From the trace decomposition we have
\be
\Tr X^2 = {\tau_2 \over 4\pi}   \left(\sum_{i} {\rm x}_i^2 + \sum_{s} {\rm x}^+_s {\rm x}^-_s \right) \,,
\ee
where we are denoting ${\rm x}_i$ as the $(K-1)+(N-K)^2$ massless fluctuations, and ${\rm x}^\pm$ the $K(2N-K-1)$ massive ones\footnote{We denote the massive modes in the upper off-diagonal part of $X$ by ${\rm x}^+$, and those in the lower off-diagonal part by ${\rm x}^-$. }, that is 
\be
\frac{2}{g_{_{\rm YM}}^2} \! \int \!d^4 x \left(\partial_\mu \overline{{\rm x}}_i \partial^\mu {\rm x}_i + \text{c.c.}\right) + \frac{2}{g_{_{\rm YM}}^2} \! \int \! d^4 x \left(\sum_\pm \partial_\mu \overline{{\rm x}}_s^\pm \partial^\mu {\rm x}_s^\pm + \frac{4(x_1-x_2)^2}{(x{-}x_1)^2(x{-}x_2)^2} M_s^2 \, {\rm x}_s^+ {\rm x}_s^- + \text{c.c.}\right) \, , 
\ee
where $M_s$ takes the values given in \eqref{eq:Ms}. 
The massless propagator connecting two massless fields is coupling independent and will contribute to $G_{\rm free}$ in \eqref{eq:factXZ}. Namely
\be
\langle \overline{{\rm x}}_i(x_1) \, {\rm x}_j  (x_2) \rangle_\theta = \frac{g_{_{\rm YM}}^2}{4\pi^2} \frac{\delta_{ij}}{x_{12}^2} \, . 
\ee
The massive propagator on the other hand will depend on the 't Hooft coupling $\lambda$ via the masses since $M_{s} \propto \sqrt{\lambda}$. Their propagator $G(x,y)$ is the solution of
\be
\left(-\Box_x +\, {4(x_1 -x_2)^2 \over (x-x_1)^2 (x-x_2)^2} M_{s}^2 \right)\, G_{s}(x,y) = \delta^{(4)} (x-y) \, .
\ee
This equation can be solved recursively order by order in the small-$\lambda$ expansion \cite{Giombi:2020enj} (see also \cite{Caetano:2023zwe}), and the solution at the order $\lambda^\ell$ is given by an $\ell$-loop conformal ladder integral.\footnote{It is interesting to note that all-loop conformal ladder integrals also appear in the study of correlation functions in other contexts \cite{Petkou:2021zhg, Karydas:2023ufs}.} This leads to the propagator of the  massive scalars, 
\be 
\begin{aligned}\label{eq:ts}
&\langle \overline{{\rm x}}_s^\pm (x_3)\, {\rm x}_s^\mp (x_4)\rangle_\theta = \frac{g_{_{\rm YM}}^2}{4\pi^2 x_{34}^2}  \sum_s t_s(u,v; \lambda) \,,  ~~~ {\rm with} ~~ t_s(u,v; \lambda) = \sum_{\ell= 0}^{\infty} (-  M_s^2 )^{\ell}\,  P^{(\ell)}(u, v) \, .
\end{aligned}
\ee
And $P^{(L)}(u, v)$ is the $L-$loop ladder Feynman integral, as described by \autoref{fig:4pt_massive} (which shows a combination of two massive propagators), 
\ie
P^{(L)}(u, v) =   \int {d^4 x_5 \over \pi^2} \ldots { d^4 x_{L+4} \over \pi^2} \frac{x_{13}^2    x_{24}^2  (x_{12}^2)^{L-1}}{ x_{45}^2  \prod_{i=5}^{L+4} x^2_{i,i+1} x^2_{1i} x^2_{2i}}\, ,
\fe
here we have identified $x_{L+5}:=x_3$. The ladder integral is known in closed form to any loop order \cite{Usyukina:1993ch}, and it is given by 
\be \label{eq:ladderL}
\begin{aligned}
&P^{(L)}(u, v) = {u \over  z -\bar{z}} \sum_{r=0}^L {(-1)^r (2L-r)! \over r! (L-r)! L! } \log^r (v) \left( \text{Li}_{2L-r}(1-z) - \text{Li}_{2L-r}(1-\bar{z}) \right) \, ,
\end{aligned}
\ee
with $P^{(0)}(u,v) = 1$. 
We then conclude  
\be
\begin{aligned}
\langle \Tr X^2(x_3) \Tr \overline{X}^2(x_4) \rangle_{{\theta}} &= 2 \left( {\tau_2 \over 4\pi } \right)^2 \sum_{i} \langle \overline{{\rm x}}_i(x_1) {\rm x}_j (x_2) \rangle_\theta^2 + 2   \left( {\tau_2 \over 4\pi } \right)^2  \sum_{s} \langle \overline{{\rm x}}^+_s(x_1) {\rm x}^-_s (x_2) \rangle_\theta^2 \\ & = 
2\frac{(K-1)+(N-K)^2}{16\pi^4x_{34}^4} + \frac{2}{16\pi^4x_{34}^4} \sum_s t^2_s(u,v; \lambda)  \,.
\end{aligned}
\ee
Note that the propagators are $\theta-$independent, therefore all $\theta-$integrals are trivial. The first term together with the $\ell=0$ term of the massive propagator conspire to give 
\be
F_6(u, v) = \frac{N^2-1}{2} \,,
\ee
consistently with equation \eqref{eq:F611}, while for the reduced correlator we finally get
\be\label{eq:Tuv_final_clas}
\mathcal{T}_{m,K}\left(u,v; \lambda \right) = \frac{1}{2 u} \sum_s \left( t_s^2(u,v;  \lambda) - 1\right)~,
\ee
where $t_s$ is given in \eqref{eq:ts}. This result is displayed in \autoref{fig:4pt_massive}, combining the perturbative diagrammatic picture with the semiclassical interpretation.

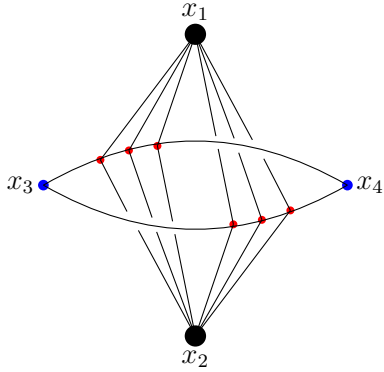
\begin{figure}[t!]
    \centering
\begin{tikzpicture}
    \coordinate (x1) at (0,2);
    \coordinate (x2) at (0,-2);
    \coordinate (x3) at (-2, 0);
    \coordinate (x4) at (2, 0);
    \coordinate (x5) at (-1.25,0.335);
    \coordinate (x6) at (-0.5,0.52);
    \coordinate (x8) at (1.25,-0.335);
    \coordinate (x7) at (0.5,-0.52);
    \coordinate (x5b) at (-0.875,0.46);
    \coordinate (x7b) at (0.875,-0.46);
    
    \fill[blue] (x3) circle (2pt);
    \fill[blue] (x4) circle (2pt);
    \fill[red] (x5) circle (1.5pt);
    \fill[red] (x5b) circle (1.5pt);
    \fill[red] (x6) circle (1.5pt);
    \fill[red] (x7) circle (1.5pt);
    \fill[red] (x7b) circle (1.5pt);
    \fill[red] (x8) circle (1.5pt);
    \fill[black] (x1) circle (4pt);
    \fill[black] (x2) circle (4pt);

    \draw[black] (x3) to[out=30, in=150] (x4);
    \draw[black] (x3) to[out=-30, in=-150] (x4);
     \draw[black] (x1) to (x5);
      \draw[black] (x1) to (x5b);
     \draw[black] (x1) to (x6);
     \draw[black] (x2) to (x7);
     \draw[black] (x2) to (x7b);
     \draw[black] (x2) to (x8);
     \draw[black] (x5) to (-0.9,-0.35);
     \draw[black] (x5b) to (-0.57,-0.43);
      \draw[black] (x6) to (-0.3,-0.48);
     \draw[black] (-0.75,-0.6) to (x2);
     \draw[black] (-0.51,-0.62) to (x2);
     \draw[black] (-0.27,-0.65) to (x2);
     \draw[black] (x8) to (0.9,0.35);
     \draw[black] (x7b) to (0.57,0.43);
     \draw[black] (x7) to (0.3,0.48);
     \draw[black] (0.51,0.62) to (x1);
      \draw[black] (0.75,0.6) to (x1);
     \draw[black] (0.27,0.65) to (x1);
     
    \node[scale=0.9] at (0,2.3) {$x_1$}; \node[scale=0.9]  at (0,-2.3) {$x_2$};
      \node[scale=0.9]  at (-2.3,0) {$x_3$}; \node[scale=0.9]  at (2.3,0) {$x_4$};
\end{tikzpicture}
    \caption{Diagrammatic representation of the four-point reduced correlator $\mathcal{T}_{m,K}$ at leading order in large charge. Each red dot represents a scalar interaction vertex, which we integrate over. In the semiclassical approach, this picture is interpreted as a combination of two massive propagators between $x_3$ and $x_4$ in the background created by the heavy canonical operators located at $x_1$ and $x_2$.}
    \label{fig:4pt_massive}
\end{figure}

As shown in \cite{Broadhurst:2010ds}, the perturbative expansion of ladder Feynman integrals can be resummed. A possible representation of the resummation can be expressed as 
\ie \label{eq:resum-form}
t_s(u,v; \lambda)= {u \over  \sqrt{v}} \sum_{r=1}^{\infty} {r \, e^{-\sigma \sqrt{r^2+4 M_s^2}} \over  \sqrt{r^2+4 M_s^2}} {\sin(r \varphi) \over \sin(\varphi)} \, ,
\fe
where $e^{i \varphi} = \sqrt{z/\bar{z}}$ and $e^{-\sigma} = \sqrt{z \bar{z}}$. Recalling the expression for $M_s^2 \sim \lambda$ as given in \eqref{eq:Ms}, we can conclude that expression \eqref{eq:resum-form} is valid non-perturbatively and applies for any values of the coupling $\lambda$.
Massaging \eqref{eq:resum-form} following \cite{Caetano:2023zwe} we can express the massive propagator at strong coupling $\lambda$ as a sum over worldline instantons. This gives
\be 
\begin{aligned}
t_s(u,v; \lambda) &= \frac{u}{\sqrt{v}} \sum_{n\ge 0} \left(W_s(\varphi+2\pi n,\sigma)+W_s(2\pi-\varphi+2\pi n,\sigma) \right) \,, \\
W_s(x,\sigma) &= \frac{2 M_s x K_1 \left(2 M_s \sqrt{x^2+\sigma^2}\right)}{\sin (x) \sqrt{x^2+\sigma^2}} \,,
\end{aligned}
\ee
where $K_1(x)$ is the modified Bessel function.

\subsection{HHLL Four-point functions: OPE analysis}\label{sec:4pt_ope}
We now discuss two different OPE limits of the four-point function discussed in the previous sections. Let us start with the \newline
{\bf{S-channel OPE:}} in this channel we fuse the heavy and the light operators together. This corresponds to the limit $z,\bar{z}\to 0$. The reduced correlators admit the S-channel  OPE expansion 
\be\label{eq:SchT}
\mathcal{T}_{m,K}(u,v; \lambda) = \sum_{\Delta,S} \sum_i \Big| C_{T_2 \mathcal{O}^m_K \mathcal{O}^{(i)}_{\Delta,S}} \Big|^2 \mathcal{G}_{\Delta,S}(u,v) \, , 
\ee
where $\{\Delta,S\}$ denote the dimension and the spin of the operator exchanged between the heavy and the light fields, and the index $i$ denotes possible degeneracies. $\mathcal{G}_{\Delta,S}(u,v)$ denotes the 4d conformal block. In the large charge limit, the heavy-light conformal block is known to simplify to a Gegenbauer polynomial $C_S^{(1)}(\varphi)$ \cite{Jafferis:2017zna}, that is
\be
\mathcal{G}_{\Delta,S}(u,v) = e^{-\sigma \left(\Delta-\Delta_H\right)} \frac{\sin (S+1)\varphi}{\sin \varphi} \,.
\ee
From the resummed expression \eqref{eq:resum-form}, following \cite{Caetano:2023zwe,Brown:2024yvt} we can rewrite the reduced correlator as 
\ie \label{eq:resummationope}
\mathcal{T}_{m,K}(u,v; \lambda) =\sum_s \sum_{S=0}^{\infty} \sum_{n=0}^{\infty} \sum^{S+1}_{r=1} C^{(s)}_{S, n, r}  \, e^{-\sigma (\Delta^{(s)}_{S, n, r}-\Delta_{\cH})} {\sin(S+1) \varphi \over \sin \varphi }\, , 
\fe
where 
\ie \label{eq:heavyOPE}
C^{(s)}_{S, n, r} &= {(r+n)(S+2+n-r) \over \sqrt{(r+n)^2+4M_s^2} \sqrt{(S+2+n-r)^2+4M_s^2}} \, , \cr 
\Delta^{(s)}_{S, n, r} &= \Delta_{\mathcal{O}_K^m} +\sqrt{(r+n)^2+4M_s^2} + \sqrt{(S+2+n-r)^2+4M_s^2} \, ,
\fe
are the OPE coefficients squared and the scaling dimension of the exchanged operator respectively, which are obtained by comparing \eqref{eq:resummationope} with \eqref{eq:SchT}. Note that at small coupling we get 
\be
\Delta^{(s)}_{S, n, r} = \Delta_{\mathcal{O}_K^m} + 2 + 2n + S + O\left(\lambda\right) \,.
\ee
At zero coupling this matches with the dimensions of heavy light double traces
\be
\mathcal{O}_K^m \Box^n \underbrace{\partial \dots \partial}_{S} T_2 \,.
\ee
{\bf T-channel OPE:} we now move to the T-channel, where we fuse the light fields together. This corresponds to the limit $u\to0$ and $v\to 1$. The T-channel OPE expansion of the reduced correlator is dominated by the Konishi exchange, that is
\be\label{eq:Tkon}
\mathcal{T}_{m,K}(u, v; \lambda) \simeq C_{\mathcal{O}_K^m {\mathcal{O}_K^m} \mathcal{O}_\mathcal{K}} C_{T_2 T_2 \mathcal{O}_\mathcal{K}} u^{\gamma_\mathcal{K}/2} - \left(C_{\mathcal{O}_K^m {\mathcal{O}_K^m} \mathcal{O}_\mathcal{K}} C_{T_2 T_2 \mathcal{O}_\mathcal{K}} \right)|_{g_{_{\rm YM}}=0} \,,
\ee
where $\gamma_\mathcal{K}$ is the anomalous dimension of the Konishi, and $C_{\mathcal{O}_K^m {\mathcal{O}_K^m} \mathcal{O}_\mathcal{K}}$, $C_{T_2 T_2 \mathcal{O}_\mathcal{K}}$ the OPE coefficients. The Konishi anomalous dimension is given by \cite{Beisert:2006ez}
\be\label{eq:kongam}
\gamma_\mathcal{K} = 3\frac{N g_{_{\rm YM}}^2}{4 \pi^2} - 3 \left(\frac{N g_{_{\rm YM}}^2}{4 \pi^2}\right)^2 + \frac{21}{4} \left(\frac{N g_{_{\rm YM}}^2}{4\pi^2}\right)^3 + \dots \, .
\ee
In terms of the large charge 't Hooft coupling, equation \eqref{eq:kongam} becomes
\be
\gamma_\mathcal{K} = 12N \frac{\lambda}{m} + O(1/m^2) \, , 
\ee
where we have only kept the leading term that will be relevant for the 't Hooft large charge limit. 
Similarly, the OPE coefficient with $T_2$ gives \cite{Eden:2012fe, Goncalves:2016vir} 
\be
\begin{aligned}
C_{T_2 T_2 \mathcal{O}_\mathcal{K}} = 1 - {3 \over 2} \frac{N g_{_{\rm YM}}^2}{4 \pi^2} + \dots  = 1 -  6N \frac{\lambda}{m}+ O(1/m^2) \, .
\end{aligned}
\ee
Finally, the OPE coefficients with the heavy operators admits the following large charge expansion \cite{Caetano:2023zwe,Brown:2024yvt}
\be
C_{\mathcal{O}_K^m {\mathcal{O}_K^m} \mathcal{O}_\mathcal{K}} = mK \, C_{\mathcal{O}_K^m {\mathcal{O}_K^m} \mathcal{O}_\mathcal{K}}^{(0)} + \sum_{L\ge1} C_{\mathcal{O}_K^m {\mathcal{O}_K^m} \mathcal{O}_\mathcal{K}}^{(L)} \lambda^L + O(1/m) \,. 
\ee
All in all, the large charge expansion of \eqref{eq:Tkon} reads
\be\label{eq:redtchope}
\mathcal{T}_{m,K}(u,v; \lambda) \simeq \sum_{L\ge1} C_{\mathcal{O}_K^m {\mathcal{O}_K^m} \mathcal{O}_\mathcal{K}}^{(L)} \lambda^L + 6N \lambda C_{\mathcal{O}_K^m {\mathcal{O}_K^m} \mathcal{O}_\mathcal{K}}^{(0)} \left(\log u-1\right) \, .
\ee
This has to be compared with the T-channel expansion of \eqref{eq:Tuv_final_clas}. In this limit we have
\be
\begin{aligned}
    P^{(1)} (u,v) &= u(2-\log u)+O(u^2,u(v-1)) \,, \\
    P^{(L\ge 2)} (u,v) &= u \binom{2L}{L} \zeta(2L-1) + O(u^2,u(v-1)) \,.
\end{aligned}
\ee
As $u\to 0$ in $t_s^2$ we only get contributions from tree level terms multiplying higher loops, as all other terms would go as $u^2$. This gives
\be
\frac{1}{2u}\left(t_s^2(u,v; \lambda)-1 \right) \simeq -M_s^2\left(2 -\log u\right) + \sum_{L\ge2} (-M_s^2)^L \binom{2L}{L} \zeta (2L-1) \,.
\ee
Finally we get for the reduced correlator
\be
\mathcal{T}_{m,K}(u,v; \lambda) \simeq 2 K N \lambda (\log u-2) + \sum_s \sum_{L\ge2} (-M_s^2)^L \binom{2L}{L} \zeta (2L-1) \, ,
\ee
where we have used the relation \eqref{eq:summssq}. 
Comparing with \eqref{eq:redtchope}, we finally obtain the OPE coefficient
\be\label{eq:Cook}
C_{\mathcal{O}_K^m {\mathcal{O}_K^m} \mathcal{O}_\mathcal{K}} \simeq \frac{mK}{3} -2 K N \lambda +  \sum_s \sum_{L\ge2} (-M_s^2)^L \binom{2L}{L} \zeta (2L-1)  \, ,
\ee
which can be resummed to 
\be\label{eq:Konoperes}
C_{\mathcal{O}_K^m {\mathcal{O}_K^m} \mathcal{O}_\mathcal{K}} \simeq \frac{mK}{3} \underbrace{-2 K N \lambda +  \sum_s  \int_0^{\infty} {M_s \, dw \over \sinh^2(w)} \left[ 2 M_s w- J_1\left(4 M_s w\right) \right]}_{C_{1}(\lambda)}\, ,
\ee
where $J_{\nu}(x)$ is the Bessel function. 
The previous expression is exact in the 't Hooft coupling. Every massive fluctuation contributes to the OPE coefficient with the integral of a Bessel $J_1$ function. Setting $K=N=2$, we get agreement with \cite{Caetano:2023zwe,Brown:2024yvt}, taking into account their normalization of the Konishi operator. As we show in \autoref{fig:COPE}, it is interesting to note that the coupling-dependent part of the OPE coefficients, namely $C_{1}(\lambda)$ in \eqref{eq:Konoperes}, has a non-trivial zero $\lambda = \lambda^*(N)$  (beyond the trivial one at $\lambda=0$).  This leads to some `non-trivial' tree level exactness of OPE coefficients 
\ie 
C_{\mathcal{O}^m_N \mathcal{O}^m_N \mathcal{O}_{\mathcal{K}}} (\lambda^*) = \frac{mK}{3} \, . 
\fe
It would be interesting to investigate how this value is modified for  higher orders in the charge expansion.

\begin{figure}[t]
\begin{center}
\begin{subfigure}[b]{0.4\textwidth}
\includegraphics[width=1.2\linewidth]{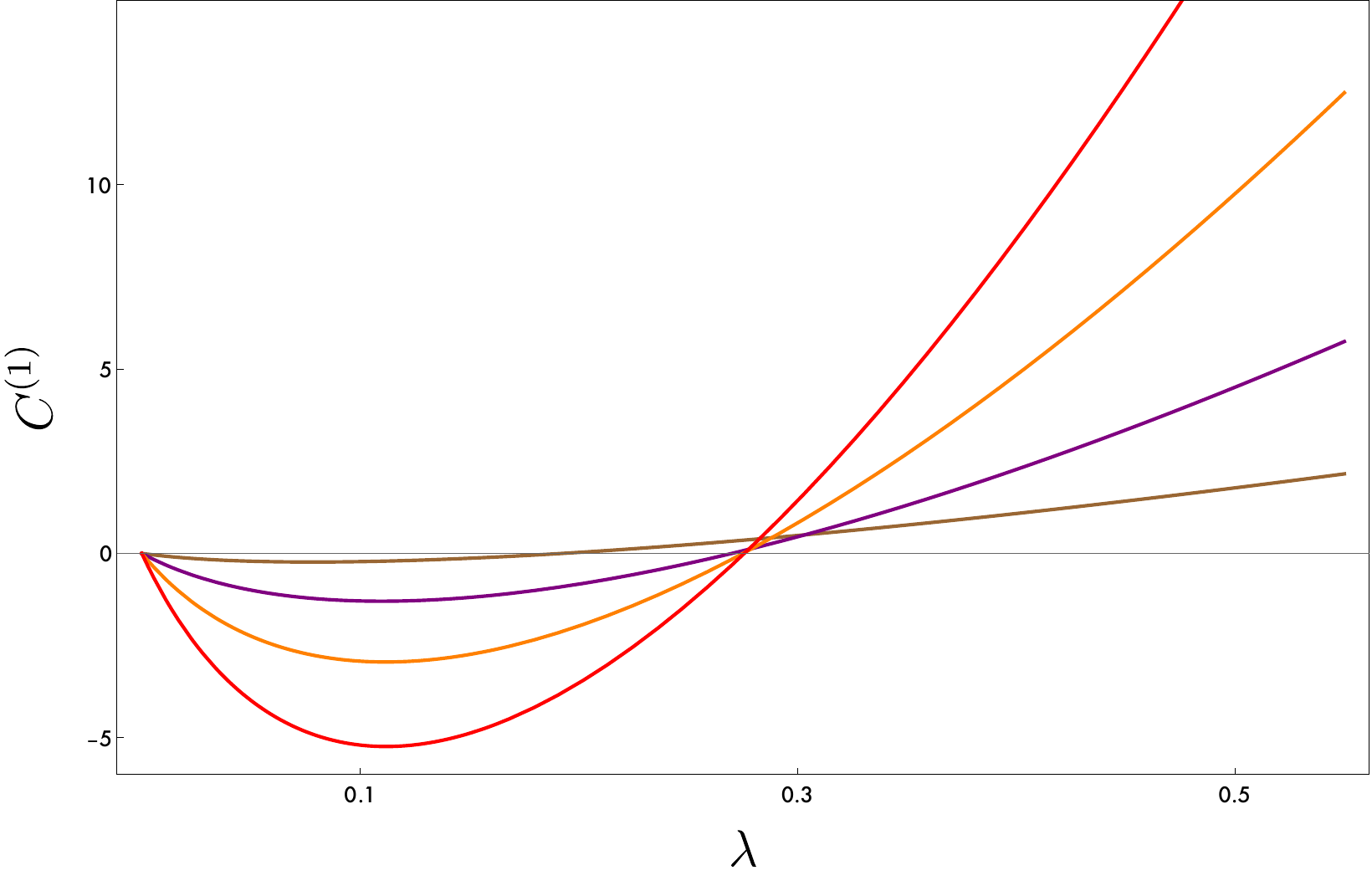}
\subcaption{}
\end{subfigure}
\hspace{2cm}
\begin{subfigure}[b]{0.4\textwidth}
\includegraphics[width=1.14\linewidth]{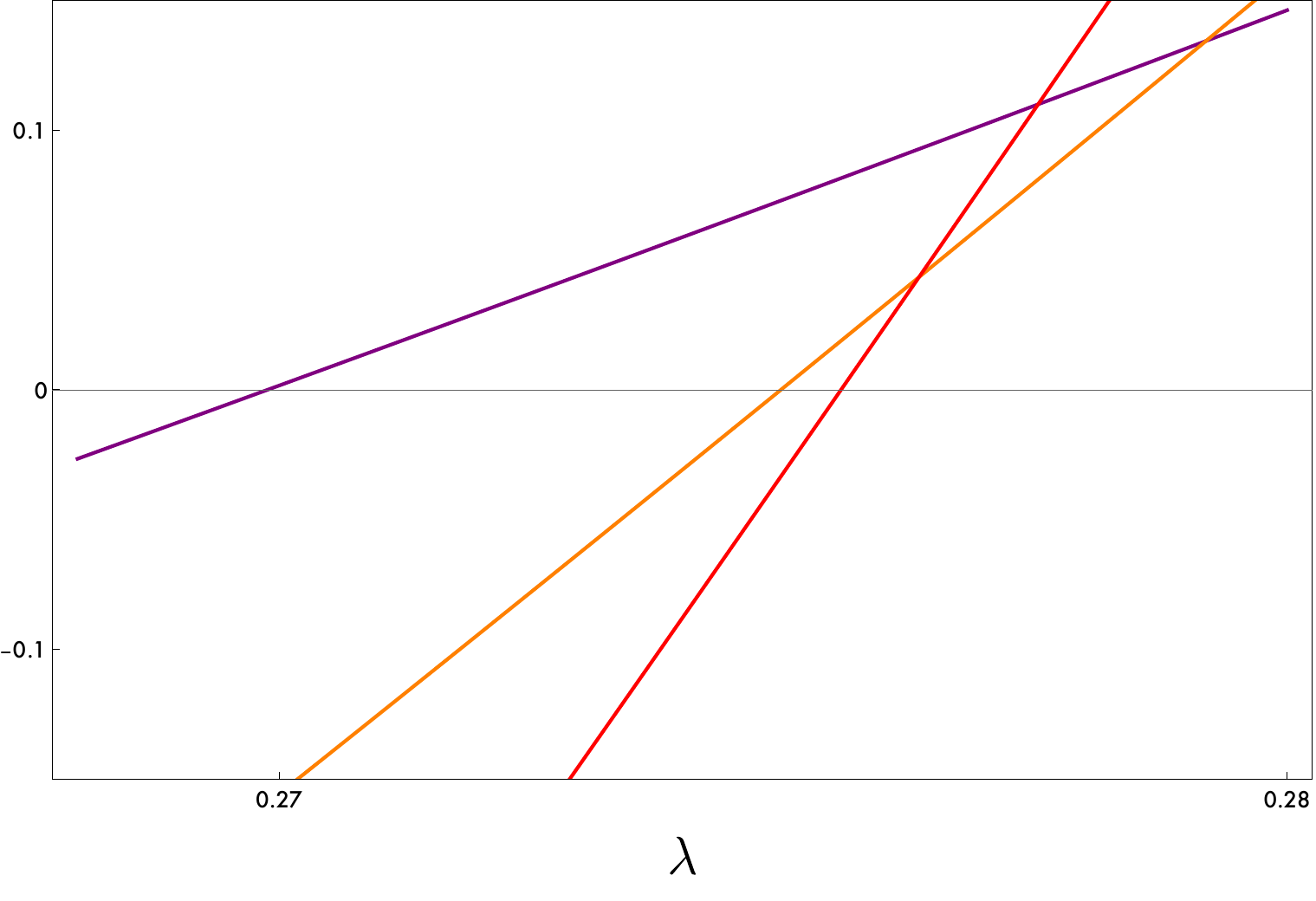}
\subcaption{}
\end{subfigure}
\end{center}
\caption{In (a) we plot the quantum correction to the OPE coefficient of the Konishi operator $C_{1}(\lambda)$ in \eqref{eq:Konoperes} for canonical operators $\mathcal{O}^m_N$ for $N =$ \textcolor{brown}{2}, \textcolor{purple}{4}, \textcolor{orange}{6}, \textcolor{red}{8}. $C_{1}(\lambda)$ vanishes at a $\lambda = \lambda^*(N)$. In (b) we zoom close to the zero of $C_{1}(\lambda)$ for $N =$ \textcolor{purple}{4}, \textcolor{orange}{6}, \textcolor{red}{8}.}
 \label{fig:COPE}
\end{figure}

\section{Supersymmetric localization and Pestun matrix model}
\label{sec:localization}

In this section, we review the basics of supersymmetric localization on  $S^4$, which we will later use in \autoref{sec:testloc}
to test our results from \autoref{sec:semi-ana}.

\subsection{The correlators via the Pestun matrix model}\label{sec:sphere}

The partition function of $\cN=4$ theory (and its deformations preserving at least $\cN=2$ supersymmetry) can be mapped to a matrix model on a four-sphere \cite{Pestun:2007rz}, where the integration variable is an $N \times N$ Hermitian matrix $a$ taking values in the $\mathfrak{su}(N)$ gauge algebra, 
\be \label{eq:matrix-mod}
a = {\rm diag} \left(a_1, a_2, \cdots, a_N\right), \qquad \sum_{i=1}^N a_i = 0~.
\ee
For the purposes of this paper, we consider the massive deformation of $\cN=4$ SYM, the so-called $\cN=2^*$ theory, whose partition function on $S^4$ can be written in terms of an integral over the eigenvalues of $a$ as follows:
\begin{equation}\label{eq:Z_matrixModel_tot}
    \begin{aligned}
\cZ(\tau, \tau_p; \mu) =\! \int \! d\sigma(a_i)  \left \vert \exp\bigg(\mathrm{i}\, \pi \tau \sum_i a_i^2 +\mathrm{i} \sum_{p>2} \pi^{p/2} \tau_p \sum_{i} a_i^p \bigg)\right\vert^2 \!   
Z_{1\textrm{-loop}}(a; \mu) \left\vert Z_{\rm inst}(\tau, \tau_p, a ; \mu) \right\vert^2\phantom{\bigg|} \, , 
\end{aligned}
\end{equation}
where $\mu$ denotes the deformation mass, and we have defined the integration measure as
\begin{equation} \label{eq:measure_a}
  d\sigma(a_i)  =  \prod_{i=1}^N da_i \,\prod_{i<j} a_{ij}^2~  \delta\bigg( \sum_{i} a_i\bigg) \, , \qquad a_{ij}:=a_{i}-a_j \, , 
\end{equation}
where the eigenvalues $a_i$ are constrained by the $\mathfrak{su}(N)$ tracelessness condition. The exponential term contains the classical action proportional to the complexified gauge coupling $\tau$, as well as the higher dimensional couplings $\tau_p,\bar\tau_p$ acting as sources for (anti-)chiral primary operators $\tr a^p = \sum_i a_i^p$ on $S^4$. $Z_{1\textrm{-loop}}$ and $Z_{\rm inst}$ correspond to the perturbative one-loop determinant and the non-perturbative instanton contributions, respectively.

For $\mu=\tau_p=0$, both $Z_{1\textrm{-loop}}$ and $ Z_{\rm inst}$ reduce to $1$ and the matrix model becomes exactly Gaussian.  In this case, the partition function reads 
\begin{equation}\label{eq:Z_matrixModelGaussian}
\cZ \equiv \cZ(\tau, 0; 0) = \int d\sigma(a_i)  \exp\bigg(\!\! -2\pi \tau_2\, \sum_i a_i^2\bigg)  \, .
\end{equation}
After rescaling the matrix $a$ as
\begin{equation}\label{eq:a_rescale}
    a \to \frac{a}{\sqrt{2\pi \tau_2}}\,,
\end{equation}
one can write down any observable $f(a_i)$ evaluated in the Gaussian matrix model as:
\begin{equation}\label{eq:f_Gaussian}
   \llangle f(a_i)\rrangle  = \frac{1}{\cZ } \int   d\sigma(a_i)  \exp\bigg(\!\! - \sum_i a_i^2\bigg) f(a_i) \, .
\end{equation}

When turning on the chiral/anti-chiral couplings $\tau_p,\bar\tau_p$, we can compute correlation functions of Coulomb Branch operators, that in the matrix model are defined as follows
\be \label{eq:ope1}
{t}_{\bs n}(a) =\prod_{k \geq 1} ( \tr a^{n_k})~, 
\ee
where $a$ is the diagonal matrix defined in \eqref{eq:matrix-mod}. 
Correlation functions of ${t}_{\bs n}$ operators are computed in the Gaussian matrix model by taking multiple derivatives with respect to $\tau_p,\bar\tau_p$. In particular their two-point function can be computed as follows
\be \label{s4corr}\ba \llangle  {{t}}_{\bs n}(a) \, { {{t}}_{\bs m}(a)} \rrangle
= {1 \over \cZ } \prod_{k \geq 1}\prod_{l \geq 1}(-\ri \pi^{1/2})^{-n_{k}}(\ri \pi^{1/2})^{-m_{l}}\partial_{\tau_{n_{k}}}\partial_{\overline \tau_{m_{l}}} 
\mathcal{Z}(\tau, \tau_{p};0)
\Big|_{\tau_{p}=0}\, .
 \ea\ee 
Correlation functions of the ${t}_{\bs n}$ operators in the matrix model on $S^4$ do not directly correspond to correlation functions of the ${T}_{\bs n}$ operators on $\mathbb R^4$. Indeed, although \(\mathbb{R}^4\) and \(S^4\) are conformally equivalent, the mapping between correlation functions on \(S^4\) and \(\mathbb{R}^4\) is highly nontrivial. This complexity arises from conformal anomalies, which lead to operator mixing on the sphere \cite{Gerchkovitz:2016gxx}: an operator of dimension $\Delta$ can mix with operators of dimensions $\Delta'=\Delta-2k$, $k>0$. To correctly address the operator mixing, it is necessary to implement an appropriate Gram-Schmidt procedure. In particular, the appropriate sphere version of the multitrace operators \eqref{eq:mul-tr} reads:
\begin{equation}\label{eq:Tn_NO}
    T_{\bs n}(a) = {t}_{\bs n}(a) +  \sum_{\bs m\,\vdash m< n}\alpha_{\bs m,\bs n}~{t}_{\bs m}(a)  \, ,
\end{equation}
where the sum in \eqref{eq:Tn_NO} runs over all the operators defined by the vector 
$\bs m = \{ m_1,\dots,m_k \}$ with dimension $m=n-2, n-4, \ldots \,$, and the symbol $\vdash$ indicates the partitions of $m$. The mixing coefficients $\alpha_{\bs m,\bs n}$ are functions of $N$ and are determined by requiring the orthogonality conditions 
\ie 
\llangle  T_{\bs n}(a) \, { {{t}}_{\bs m}(a)} \rrangle=0 \, ,  \qquad \bs m\,\vdash m< n \, . 
\fe
In the next subsection we will discuss the derivation of the class of canonical operators on the sphere, and we will provide explicit examples for the mixing coefficients.
The Gaussian matrix model \eqref{eq:f_Gaussian} together with the Gram-Schmidt procedure \eqref{eq:Tn_NO} are the main tools to compute the two-point normalization $\cN_{\bf{m}}$ and the three-point coefficient $\mathfrak{C}_{\bf{m}\bf{m}\bf{p}}$ in flat space.

\subsubsection{Integrated correlators via the matrix model}
Considering the HHLL four-point correlators of \eqref{eq:4ptHH22}, the reduced correlator $\cT_{m,K} (u,v; \tau, \bar{\tau})$ contains an explicit spacetime dependence through the conformal cross-ratios $u,v$ and cannot directly be computed by the matrix model. However, we can define the integrated correlators by integrating out the spacetime dependence with a special measure, which once again can be obtained from the matrix model \cite{Binder:2019jwn, Chester:2020dja}. This class of observables has been studied extensively in the literature in recent years.  Notably, they can be computed exactly as functions of the Yang-Mills coupling $\tau$ by leveraging their modular structures underlying the S-duality of $\mathcal{N}=4$ SYM and  mathematical properties of (non-holomorphic) modular forms \cite{Chester:2019jas, Chester:2020vyz, Dorigoni:2021bvj, Dorigoni:2021guq, Collier:2022emf, Paul:2022piq, Dorigoni:2022zcr, Dorigoni:2025beu}.

Considering the reduced correlator  $\cT_{m,K} (u,v; \tau, \bar{\tau})$, the corresponding integrated correlator is defined as: 
\begin{equation} \label{eq:def-integration}
\mathcal{G}_{m, K}(\tau,\bar\tau)=I_2\left[\cT_{m,K} (u,v; \tau, \bar{\tau}) \right]=-\frac{8}{\pi} \int_0^{\infty} dr \int_0^{\pi}  d\theta \frac{r^3 \sin^2 \theta}{u \, v}  \cT_{m,K} (u,v; \tau, \bar{\tau})~.
\end{equation}
The integrated correlators are non-trivial functions of the coupling $\tau$ as made clear in \eqref{eq:def-integration}. In order to compute them using the matrix model, we now need to consider the mass-deformed matrix model as in \eqref{eq:Z_matrixModel_tot}. 
In that case, the one-loop determinant reads:
\begin{equation}\label{eq:Z1loop}
    Z_{1\textrm{-loop}}(a;\mu) = \frac{1}{H(\mu)^N} \prod_{i<j} \frac{H^2(a_{ij})}{H(a_{ij}+\mu) H(a_{ij}-\mu)}\,,
\end{equation}
where $H(x)$ is a product of Barnes G-functions $G(x)$:
\begin{equation}\label{eq:barnes}
    H(x) = \mathrm{e}^{-(1+\gamma)x^2} G(1+\mathrm{i} x) G(1-\mathrm{i} x)\,.
\end{equation}
For the purpose of this work we can consider the zero-instanton sector and set $Z_{\mathrm{inst}}=1$.\footnote{This is justified by the fact that all the Yang-Mills instanton contributions in the large-charge 't Hooft limit \eqref{eq:thin} are exponentially suppressed as \(O(\re^{-m/\lambda})\), contributing only non-perturbatively.}  The main result of \cite{Binder:2019jwn} states that the integrated correlators defined in \eqref{eq:def-integration} can be obtained by taking appropriate derivatives on the $\mathcal{N}=2^*$ partition function in \eqref{eq:Z_matrixModel_tot}.  For the correlators we consider here, we have: 
\begin{equation}\label{eq:22mm_example}
    \cG_{m, K}(\tau,\bar\tau) =    \dfrac{\displaystyle  \,\partial_{\tau_{m,K}} \partial_{\bar\tau_{m,K}} \partial_{\mu}^2 \log \cZ(\tau, \tau_p; \mu) \, {\vert}_{\tau_p,\, \mu=0}}{\displaystyle \partial_{\tau_{m,K}} \partial_{\bar\tau_{m,K}} \log \cZ(\tau, \tau_p; 0) \, {\vert}_{\tau_p=0}}\,,
\end{equation}
where $\partial_{\tau_{m,K}}$ (and $\partial_{\bar\tau_{m,K}}$) schematically represent the insertions of two canonical operators $\cO_{K}^m$, their explicit construction will become clear shortly in the next subsection. The two mass derivatives $\partial_{\mu}^2$ realize the insertion of the $T_2$'s\footnote{In deriving the localization formula for the integrated correlators, one needs the partial supersymmetry breaking to $\mathcal{N}=2^*$ theory. In this set up, the operators $T_2$ are coupled to the deformation mass $\mu$ and are inserted by taking derivatives with respect to the mass.}.  For carrying out the matrix model calculation explicitly, we will need the $\mu^2$ term  (the first nontrivial mass correction)  of the one-loop determinant \eqref{eq:Z1loop}. After implementing the coupling rescaling \eqref{eq:a_rescale}, it can be expressed as follows:
\begin{align}\label{eq:z1loop_pert}
\log Z_{\textrm{1-loop}}=-\mu^2 \bigg[\sum_{L=1}^\infty\sum_{j=0}^{2L}(-1)^{L+j}\left(\frac{1}{2\pi\tau_2}\right)^L\binom{2L}{j} (2L {+} 1) \, \zeta(2L {+} 1) \, \tr\,a^{2L-j} \tr\,a^{j}\bigg]+O(\mu^4)\, .
\end{align} 
This expression will be relevant for computing the integrated four-point function in the following section.

\subsection{Canonical operators on $S^4$}\label{sec:canonical}

We now compute canonical operators using the $S^4$ formalism, following \cite[Sec.~3.2.1]{Gerchkovitz:2016gxx}. A similar construction has been used in \cite{Brown:2023cpz, Brown:2023why, Grassi:2024bwl} in the computation of integrated correlators in presence of different large charge operators.  When transitioning back from $S^4$ to $\IR^4$, this construction provides the coefficients for the orthogonal basis of superconformal primaries given in  \autoref{sec:definition}.

It is convenient to introduce a specific ordering, as it ensures a systematic computation of the correlators. Other orderings can be related through a change of basis. 
Operators of the same dimension are ordered as follows: operators with the fewest factors of $(\tr  a^2)$ come first, and if these are the same, then the operators with the fewest factors of $(\tr  a^3)$ come first, and so on up to $(\tr  a^N)$. For example, $\Delta = 12$, $N=5$ will be ordered as
\ie \label{eq:example}
\{t_{4,4,4}, \, t_{5,4,3}, \, t_{3,3,3,3}, \, t_{5,5,2},  \, t_{4,3,3,2}, \, t_{4,4,2,2}, \, t_{5,3,2,2}, \, t_{3,3,2,2,2}, \, t_{4,2,2,2,2}, \, t_{2,2,2,2,2,2}\}\, . 
\fe
We then define the new operators 
\begin{equation}
\mathcal{O}_{\bf {m}}(a), \qquad {{\bf m}} = \{m_1, m_{2}, \dots \} \, ,
\end{equation}
by applying Gram-Schmidt orthogonalization to $t_{\textbf{m}}$ with all operators after it (i.e. to its right). Note that this does not always mean that operators with fewer traces come before operators with higher numbers of traces. In the Gram-Schmidt process, the scalar product is defined by using the Gaussian measure \eqref{eq:f_Gaussian}.

As the last step, we define  the following set of operators
\begin{equation}
\left\{ t_{2}^n \mathcal{O}_{\bf {m}}(a)\right\}_{n,{\bf{m}}\geq 0}~.
\end{equation}
This set of operators forms a basis of the chiral ring with the property that
\be \llangle t_{2}^n \mathcal{O}_{\bf {m}}(a), t_{2}^{n'} \mathcal{O}_{\bf {m'}}(a) \rrangle \propto \delta_{\bf {m}, \bf {m'} } \, . \ee
In particular \( \mathcal{O}_{\bf {m}} (a)\) are orthogonal to all the operators of equal or smaller dimension, hence they do not undergo operator mixing when going from $S^4$ to $\IR^4$. As a consequence,
their two point functions on $\IR^4$ and on $S^4$ are equivalent.

We then concentrate on the $\mathcal{O}_{\bf m} $ operators in the large-charge limit. 
In this regime, all operators of the form 
\be
\mathcal{O}_{i_1, \dots i_{k-1}, 
\begin{tikzpicture}[baseline=(O.base)]
    \node (O) {$i_k, \dots, i_k$};
    \draw [decorate,decoration={brace,mirror,amplitude=3pt,raise=1pt}] 
        (O.south west) -- (O.south east) node[midway,below=3pt] {\footnotesize $m$};
\end{tikzpicture},
 i_{k+1}, \dots i_n}
\quad\text{with}~ ~i_\ell\neq i_j~~ \text{if} ~~ \ell\neq j \, ,
\ee
exhibit the same asymptotic behavior  as $m\to \infty$. Specifically,  we find that within correlation functions:
\be \label{eq:exampsub}
\mathcal{O}_{i_1, \dots i_{k-1}, 
\begin{tikzpicture}[baseline=(O.base)]
    \node (O) {$i_k, \dots, i_k$};
    \draw [decorate,decoration={brace,mirror,amplitude=3pt,raise=1pt}] 
        (O.south west) -- (O.south east) node[midway,below=3pt] {\footnotesize $m$};
\end{tikzpicture},
 i_{k+1}, \dots i_n}
\simeq 
\mathcal{O}_{
\begin{tikzpicture}[baseline=(O.base)]
    \node (O) {\!\!\! $i_k, \dots, i_k$};
    \draw [decorate,decoration={brace,mirror,amplitude=2pt,raise=1pt}] 
        (O.south west) -- (O.south east) node[midway,below=2pt] {\footnotesize $m$};
\end{tikzpicture}}.
\ee   
where $\simeq $ indicates equality at leading order in the large-charge (double-scaling) limit of the correlators.

This leads us to focus on operators of the form $\cO_{K,\dots,K}$, 
which we denote as 
\begin{equation} \label{Omk} 
\mathcal{O}^m_K(a)= 
t_{\begin{tikzpicture}[baseline=(O.base)]
    \node (O) {\!\!\!\! \footnotesize{$K, \dots, K$}};
    \draw [decorate,decoration={brace,mirror,amplitude=3pt,raise=1pt}] 
        (O.south west) -- (O.south east) node[midway,below=3pt] {\footnotesize $m$};
\end{tikzpicture}}\!\! (a)
+  \sum_{{\bf n}} c_{\bf n} t_{\bf{n}}(a)~,
\end{equation}
where $K \leq N$ for $SU(N)$
and the coefficients $c_{\bf n}$  are determined by the GS procedure outlined previously. Some examples are provided below. 
Among the operators \eqref{Omk}, there is one type which exhibits remarkably simple behavior, not only at leading order but also at all orders in the large charge limit. These operators are the {\it maximally symmetry breaking  operators}, specifically the operators \eqref{Omk} with $K=N$. We will discuss their special behaviour in \autoref{sec:2pnt}.

It is important to note that one can simply replace $t_{\bf p}$ in \eqref{Omk} by $T_{\bf p}$ defined in \eqref{eq:Tn_NO}, {\it i.e.} the ones that can be directly identified with the flat-space operators. This is because $\mathcal{O}_{\bf m}$ by construction is orthogonal to all the lower-dimensional operators, and $T_{\bf p}$ has exactly the same properties. All the lower-dimensional operators appear in the definition of $T_{\bf p}$ cancel out automatically in $\mathcal{O}_{\bf m}$. Therefore, $\mathcal{O}_{\bf m}$ constructed in this way should be identified with the orthogonal basis we introduced in \autoref{sec:definition} for the field theory in flat space.

Let us give some concrete examples.

\subsubsection{The example of $SU(4)$}\label{sec:egsu4o}
Let us consider the example of $SU(4)$ and 
 the operators $\mathcal{O}_{\bf m}$  at a given dimension $\Delta=3 m_3+4m_4$. We get\footnote{We omit the $a$ dependence for the sake of simplicity in notation.}
  \allowdisplaybreaks{
\begin{align}\label{eq:opsu4}
    &\mathcal{O}_{\underbrace{ 4,\dots,4}_{m_4=\Delta/4}}  = t_4^{m_4} + \text{GS}\, ~\, {\rm with} \quad \{ t_4^{n_4} t_3^{n_3} \}_{n_3\ge1} \cup \{ t_4^{n_4} t_3^{n_3} t_2^{n_2} \}_{n_2\ge1} \,, \cr
    &\vdots \cr
&\mathcal{O}_{\underbrace{4,\dots,4}_{m_4},\underbrace{3, \dots ,3}_{m_3}}  = t_4^{m_4} t_3^{m_3} + \text{GS}\, ~\,  {\rm with} \quad \{ t_4^{n_4} t_3^{n_3} \}_{n_3\ge m_3+1} \cup \{ t_4^{n_4} t_3^{n_3} t_2^{n_2} \}_{n_2\ge1} \,, \\
    &\vdots \cr
&\mathcal{O}_{\underbrace{3,\dots,3}_{m_3=\Delta/3}}  = t_3^{m_3} + \text{GS}\, ~\,    {\rm with} \quad \{ t_4^{n_4} t_3^{n_3} t_2^{n_2} \}_{n_2\ge1} \,. \nonumber
\end{align}} 
Here we list a few explicit examples. The canonical operator at $K=3, m=5$ is $\mathcal{O}_3^5=\mathcal{O}_{3,3,3,3,3} $  and reads
\be  \label{eq:exsu4_a}
\mathcal{O}^5_3 = t_3^5 
+  \frac{35325}{39442} t_3^3 t_2^3 - \frac{90}{41} t_4 t_3^3 t_2 
+ \frac{1050327}{9781616} t_3 t_2^6 - \frac{22761 }{39442} t_4 t_3 t_2^4 + \frac{405}{533}t_4^2 t_3 t_2^2 \phantom{\Bigg|}~.
\ee
Likewise the canonical operator at $K=4, m=3$ is $\mathcal{O}_4^3=\mathcal{O}_{4,4,4} $  and reads
\be \label{eq:exsu4_b}
\mathcal{O}_4^3 = \, t_4^3 
- \frac{6103 }{5140} t_4^2 t_2^2
+ \frac{9551}{20560} t_4 t_2^4 - \frac{162}{1285} t_4 t_3^2  t_2
+ \frac{53}{6939} t_3^4
+  \frac{1313}{23130} t_3^2 t_2^3 
- \frac{985 }{16448}t_2^6 ~.\phantom{\Bigg|}
\ee
As commented above, thanks to this special construction we are allowed to replace all the $t_{\bf p}$ in the above expressions by the normal-ordered $T_{\bf p}$ defined in \eqref{eq:Tn_NO}, which can be directly identified with the flat-space operators.

\subsubsection{The example of $SU(5)$}

For $SU(5)$, the operators with dimension $\Delta = 5m_5+4m_4+3m_3$ are given by
\begin{align}\label{eq:opsu5}
\mathcal{O}_{\underbrace{5,\dots,5}_{m_5},\underbrace{4, \dots ,4}_{m_4},\underbrace{3, \dots ,3}_{m_3}}  = &\, t_5^{m_5}t_4^{m_4} t_3^{m_3}\, + \\[-.4cm] &\text{GS}\, ~\,  {\rm with} ~~ \{t_5^{n_5} t_4^{n_4} t_3^{m_3} \}_{n_4\ge m_4+1} \cup \{ t_5^{n_5} t_4^{n_4} t_3^{n_3} \}_{n_3\ge m_3+1} \cup \{t_5^{n_5} t_4^{n_4} t_3^{n_3} t_2^{n_2} \}_{n_2\ge1} \,,\notag
\end{align}
and so the canonical operators are of the form
\begin{equation}
\begin{aligned}
    &\mathcal{O}_5^m = t_5^m + \text{GS}\, ~ \,{\rm with} \quad \{ t_5^{n_5} t_4^{n_4} t_3^{n_3} \}_{n_3+n_4\ge1} \cup \{  t_5^{n_5} t_4^{n_4} t_3^{n_3} t_2^{n_2} \}_{n_2\ge1} \,, \\
    &\mathcal{O}_4^m  = t_4^m + \text{GS}\, ~ \, {\rm with} \quad \{ t_5^{n_5} t_4^{n_4} t_3^{n_3} \}_{n_3\ge1} \cup \{  t_5^{n_5} t_4^{n_4} t_3^{n_3} t_2^{n_2} \}_{n_2\ge1} \,, \\
    &\mathcal{O}_3^m  = t_3^m + \text{GS}\, ~ \,  {\rm with} \quad  \{  t_5^{n_5} t_4^{n_4} t_3^{n_3} t_2^{n_2} \}_{n_2\ge1} \,. \\
\end{aligned}
\end{equation}
For example the  maximally symmetry breaking  operator is
\be \label{eq:exsu5_a} \mathcal{O}_5^2=
t_5^2 - \frac{1}{90} t_4 t_3^2 - \frac{23}{800} t_4^2  t_2 - \frac{443}{300} t_5 t_3 t_2 + \frac{2461 }{4500}t_3^2 t_2^2 + \frac{179}{8000} t_4 t_2^3- \frac{7}{1600} t_2^5~.
\ee
From the matrix model computations, as written in the examples \eqref{eq:exsu4_a}, \eqref{eq:exsu4_b} and \eqref{eq:exsu5_a}, we can read the explicit expressions of the coefficients $c_{\bs n}$ for the chiral ring basis defined in \autoref{sec:definition} on flat space, see around eq. \eqref{eq:ex55}.

\section{Correlation functions via localization} \label{sec:testloc}
In this section we provide strong evidence of the results for the correlation functions in the large-charge 't Hooft limit obtained from the semiclassical analysis in \autoref{sec:semi-ana}. We explicitly compute these correlators using the matrix model arising from supersymmetric localization as described in \autoref{sec:localization}. In particular the  two- and three-point functions of half-BPS operators are tree-level exact, thus we use the matrix model as a tool for the exact computations of the normalization factor $\cN_{m}$ and the three-point coefficient $\mathfrak{C}_{mm \bs p}$, defined in \eqref{eq:2pt_pp} and \eqref{eq:3pt_HHp} respectively. When considering the large-charge 't Hooft limit, supersymmetric localization also allows us to compute the full four-point HHLL correlator. Indeed,  Pestun's matrix model is the main technique for the computation of integrated four-point functions, where the spacetime dependence of the correlators is integrated over a supersymmetric invariant measure \cite{Binder:2019jwn}. As shown in \cite{Brown:2024yvt} and reviewed in \autoref{sec:largecharge}, the four-point correlators \eqref{eq:4ptHH22} in the large-charge 't Hooft limit are fully determined by the corresponding limit on the integrated correlators. Hence localization results will represent a direct check of the semiclassical analysis from \autoref{sec:semi-ana}.

\subsection{Two-point functions}\label{sec:2pnt}

The two-point functions of \( \frac{1}{2} \)-BPS operators are defined in \eqref{eq:2pt_pp}. Since the canonical operators on the sphere are orthogonal to all operators of the same or lower dimension,  the coefficient
$\mathcal{N}_m(N,K)$ is simply given by the two-point function on the sphere 
\be \mathcal{N}_m(N,K)=\llangle\mathcal{O}_K^m(a) \mathcal{O}_K^m(a)\rrangle \, ,
\ee
where the expectation value is taken as in \eqref{eq:f_Gaussian}. We detail the techniques used to perform explicit computations in the Gaussian matrix model in \autoref{app:Gauss_MM}.

Let us first consider the special canonical operators with $K=N$. 
For case $N=3$, it was observed in \cite{Grassi:2024bwl} that the two-point function of such an operator exhibits a behavior that mimics the EFT predictions for rank-one theories \cite{Hellerman:2018xpi,Hellerman:2017sur} at all orders in $1/m$. Notably, this holds for all canonical operators $\mathcal{O}_N^m$. We find
\be\label{eq:tpoN}  \mathcal{N}_m(N,N)= f_{N}(m) \left(2^{-N} N^{2-N} \right)^m\, \frac{\Gamma\left(N m +\frac{N(N-1)}{2}+1\right)}{\Gamma\left(\frac{N(N-1)}{2}+1\right)}\, , \ee
where for the first few $N$ we have
 \allowdisplaybreaks{ \begin{equation}\label{eq:fexa}
\ba
f_{2}(m) &= 1 \,, \\
f_{3}(m) &= 1 \,, \\
f_{4}(m) &= \frac{1-2^{-4}}{1-2^{-4-4m}}=\frac{15}{16}+ {O}(\re^{-m}) \,, \\
f_{5}(m) &= \frac{\frac{504}{625}}{\left( 1 + 5^{-\frac{5}{2} m - 4} \left( 1 + (-1)^m \right) - 2^{-m - 2} \, 5^{-3m - 5} \left( (25 + 11 \sqrt{5})^{m+2} + (25 - 11 \sqrt{5})^{m+2} \right) \right)} \\
&= {504\over 625} + {O}(\re^{-m})\, . 
\ea
\end{equation}}
\hspace{-0.3cm} We have verified the above results up to $m=70$ for $SU(4)$ and up to $m=20$ for $SU(5)$.
In particular, \eqref{eq:tpoN} matches the EFT prediction of \cite{Hellerman:2018xpi,Hellerman:2017sur} for rank-$1$ theories at all orders in the $1/m$ expansion. Indeed the $\Gamma$ function in \eqref{eq:tpoN} can be written as
\be \label{gammabeh} \Gamma\left({R\over 2}+\alpha_{N,N}+1\right)\, , \ee
where  $R= 2 N m$ is the R-charge of the $\mathcal{O}^{m}_N$ operator and $\alpha_{N,N}={N-1\over 2}N$ is precisely  the Wess-Zumino coefficient for the $a$-anomaly in $\mathcal{N}=4$ SYM (see \cite[(A.32)]{Hellerman:2017sur}). Therefore, for the canonical operators with $K=N$, not only are our expectations from \autoref{sec:2pteft} confirmed, but the analogy with \cite{Hellerman:2018xpi, Hellerman:2017sur} remains valid at all orders in the large charge expansion.

We now consider canonical operators with $K<N$.
We find that the large-charge behavior of the two-point functions of \(\mathcal{O} ^{m}_K\), is similar to that of rank-1 theories \eqref{gammabeh}, but only for the first few terms in the large $m$ expansion. 
In particular, we correctly reproduce the behavior predicted in \eqref{eq:log2pn}, but the subleading orders $\mathcal{O}\left({1\over mK}\right)$ do not match a $\Gamma$ function behavior. 
As an example we can consider $SU(4)$ with $K=3$.
By computing the correlators for many values of $m$ (we computed  correlators up to $m=72$),  we find numerically that 
\be \ba  \label{eq:o303}\llangle\mathcal{O}^m_3(a)\mathcal{O}^m_3(a)\rrangle =24^{-m} \frac{\Gamma(3m+7)}{\Gamma(7)}\left(\frac{20}{27}+\frac{250}{243m}+\mathcal{O}\left({1\over m^2}\right)\right) \, .
\ea\ee
Likewise for  $SU(5)$ with $K=3$ we get
\be \ba  \label{eq:o4}
\llangle\mathcal{O}^m_3(a)\mathcal{O}^m_3(a)\rrangle= 24^{-m} \frac{ \Gamma (3 m+10)}{\Gamma (10)}\left(\frac{140}{243}+\mathcal{O}\left({1\over m}\right)\right) \, .
\ea\ee
We generally find that $\mathcal{N}_m(N,K)$ is of the form
\begin{equation}
   \mathcal{N}_m(N,K) = \left(2^K K^{K-2} \right)^{-m} \frac{\Gamma(\Delta+\alpha_{N,K}+1)}{\Gamma(\alpha_{N,K}+1)} A(N,K) \left(1+\mathcal{O}\left({1\over m}\right)\right) \, .
\end{equation}
Expanding $\log \mathcal{N}_m(N,K)$ at large $m$, we recover equation \eqref{eq:log2pn}.

\subsection{Three-point functions}\label{sec:3pnt}

Let us now consider the three-point correlators \eqref{eq:3pt_HHp} involving two canonical operators and one superconformal primary operator. We recall that the three-point correlator takes the following form 
\begin{align}\label{eq:3pt_2}
\frac{\vev{\mathcal{O}^m_K(x_1,Y_1) \mathcal{O}^m_K(x_2,Y_2) T_{\bf p}(x_3,Y_3)} }{\vev{\mathcal{O}^m_K(x_1,Y_1) \mathcal{O}^m_K(x_2,Y_2)}}= \mathfrak{C}_{m m {\bf p}}(N,K)\times \left(\frac{1}{2\pi^2}\frac{d_{23}d_{31}}{d_{12}}\right)^{\frac{\Delta_{\bf p}}{2}}\,,
 \end{align}
where the coefficients $\mathfrak{C}_{m m {\bf p}}$ can be computed by free-theory Wick contractions. Therefore, they can effectively be obtained via the Gaussian matrix model \eqref{eq:f_Gaussian} as\footnote{Similar computations for normalized BPS three-point function via the Gaussian matrix model were performed in \cite{Brown:2024tru} in presence of determinant operators.}
\ie
\mathfrak{C}_{m m {\bf p}}(N,K) = \frac{\llangle\mathcal{O}^m_K(a) \mathcal{O}^m_K(a) T_{\bf p}(a)\rrangle }{\llangle\mathcal{O}^m_K(a) \mathcal{O}^m_K(a)\rrangle} \, , 
\fe
which can be evaluated explicitly by using the techniques outlined in \autoref{app:Gauss_MM}.
We aim to study these correlators in the large-charge limit $m \to \infty$, with the dimension of the superconformal primary $\Delta_{\bf p}$ held fixed. 

As a first observation, we note that the contribution from mixing of the operator \(T_{\mathbf{p}}\) with respect to lower dimensional operators is subleading in this limit. That is:  
\be \label{eq:3pst1}\frac{\llangle\mathcal{O}^m_K(a) \mathcal{O}^m_K(a) T_{\bf p}(a)\rrangle }{\llangle\mathcal{O}^m_K(a) \mathcal{O}^m_K(a)\rrangle} = \frac{\llangle\mathcal{O}^m_K(a) \mathcal{O}^m_K(a) t_{\bf p}(a)\rrangle }{\llangle\mathcal{O}^m_K(a) \mathcal{O}^m_K(a)\rrangle}\left(1+O\left({1\over m}\right)\right) \, .\ee
To see this, as discussed in \eqref{eq:Tn_NO}, the difference between $T_{ \bf p}$ and $t_{ \bf p}$ is the operators with lower dimensions than $\Delta_{\bf p}$, whose contributions to the three-point function are subleading in the large-charge limit.
We can then perform several explicit computations using the Pestun matrix model as outlined in \autoref{sec:sphere}. We have done these computations for all these three-point functions with $N<8$.  Some examples are provided in \autoref{App:threept}. As a result, we see that the matrix model three-point coefficients can be expressed as:
\begin{equation}\label{eq:3pres}
\frac{\llangle\mathcal{O}^m_K(a) \mathcal{O}^m_K(a) t_{\bf p}(a)\rrangle }{\llangle\mathcal{O}^m_K(a) \mathcal{O}^m_K(a)\rrangle} =  \int ^{2\pi}_0 \frac {d \theta}{2\pi} \, \langle t_{\bf p}  \rangle_\theta \left(1+ O\left({1\over m}\right)\right) \,,
\end{equation}
where
 \begin{equation}\label{eq:acl}
\langle t_{\bf p}  \rangle_\theta =  \prod_{k\geq 1} \left(
\text{Tr} \left(a_{\rm cl}^{p_k} (\theta) \right) 
\right) \,,\quad {\rm and} \quad a_{\rm cl}(\theta)={\Omega_K^{(N)}\rm e^{\rm i \theta}  +\overline{\Omega}_K^{(N)}\rm e^{-\rm i \theta} \over \sqrt{2}} \, ~.
\end{equation}
Thus the result \eqref{eq:3pres} can be interpreted as the matrix model realization of the semiclassical analysis from \autoref{sec:semi_3pt}.

\subsection{Four-point functions}\label{sec:4pt_localis}
We now analyze the class of four-point correlators discussed in \autoref{sec:4pt_interact} using semi-classical analysis. More specifically, we re-derive the result \eqref{eq:Tuv_final_clas} for the reduced four-point correlator from the approach of \cite{Brown:2024yvt}, which combines the Feynman diagrammatic analysis in the large-charge 't Hooft limit with the supersymmetric localization calculation for the integrated correlators.

\subsubsection{Review: HHLL correlators from integrated correlators}\label{sec:largecharge}

We first review the diagrammatic argument from \cite{Brown:2024yvt} for general $\langle \mathcal{H} \mathcal{H} T_2 T_2 \rangle$ correlators in $\mathcal{N}=4$ SYM valid for an $SU(N)$ gauge group, where $\mathcal{H}$ can be any half-BPS superconformal primary operators with large conformal dimension $\Delta_{\mathcal{H}}$ such that $\Delta_{\mathcal{H}}\gg N^2$. It was argued in \cite{Brown:2024yvt} that the spacetime dependent part of the correlator at a given loop order is completely fixed when considering the large-charge 't Hooft double scaling limit given in \eqref{eq:thin}. 
In particular, employing large-charge combinatorial arguments on the chiral Lagrangian insertions \cite{Eden:2012tu} along with SUSY non-renormalization theorems \cite{Baggio:2012rr}, we identify the Feynman integrals contributing at the leading order in the 't Hooft large-charge limit. In full generality, one can show that the dynamical part of the correlators $\langle \mathcal{H} \mathcal{H} T_2 T_2 \rangle$, i.e. the reduced correlator $\mathcal{T}_{\mathcal{H}}(u, v; \lambda)$ following the decomposition \eqref{eq:4ptHH22}, can be computed in terms of the following expression to all loops in perturbation theory:
\begin{equation}\label{eq:Tuv_gen}
    \mathcal T_{\mathcal{H}}(u, v; \lambda) =  \sum_{L=1}^{\infty} d_{\mathcal{H}, N; L} \,  \frac{(-\lambda)^L}{ u }\sum_{\ell=0}^{L}   P^{(\ell)}(u, v)  P^{(L-\ell)}(u, v) \, ,
\end{equation}
where $P^{(\ell)}(u, v)$ is the $\ell$-loop ladder Feynman integral as defined in \eqref{eq:ladderL}. We remark that the spacetime dependent part of \eqref{eq:Tuv_gen} is universal, and the only dependence on the precise form of the heavy operator $\cH$ lies in the color factors $d_{\mathcal{H}, N; L}$ at $L$-loop order.  As emphasized in \cite{Brown:2024yvt}, for the cases $d_{\mathcal{H}, N; L} \sim c^L$ for some constant $c$, the perturbative expression \eqref{eq:Tuv_gen} can be resummed and is valid non-perturbatively. We will return to this point when discussing the four-point function that involves canonical operators introduced in \autoref{sec:definition}. 

An efficient way to determine the color factors $d_{\mathcal{H}, N; L}$ is to calculate the integrated HHLL correlators as defined in \eqref{eq:def-integration}.  
Applying the $(u,v)$-integral to the general expression given in \eqref{eq:Tuv_gen} and using the following result \cite{Dorigoni:2021guq}
\begin{equation}\label{eq:ladder_integration}
    I_2\left[ \frac{1}{u}\sum_{\ell=0}^{L}   P^{(\ell)}(u, v)  P^{(L-\ell)}(u, v)\right] 
    =-4{\Gamma(2L+2) \over \Gamma(L+1)^2}\, \zeta (2 L+1) \, ,
\end{equation}
we then find the integrated HHLL correlators can be written as 
\begin{equation} \label{eq:gH-Feyn}
    \mathcal{G}_\cH (\lambda)  = -4\sum_{L=1}^{\infty} d_{\cH, N; L}\, (-\lambda)^L\,  {\Gamma(2L+2) \over \Gamma(L+1)^2} \zeta (2 L+1) \, . 
\end{equation}
As we discussed in \eqref{eq:22mm_example}, the integrated correlators can be computed independently using supersymmetric localization. Therefore, knowing the integrated correlators from localization allows us to deduce the color coefficients $d_{\cH, N; L}$ \cite{Brown:2024yvt}. In this way, from the integrated correlators obtained via matrix model computations, we determine the full HHLL correlators by substituting the results of $d_{\cH, N; L}$ into \eqref{eq:Tuv_gen}.

We will now apply this procedure to the HHLL correlators with canonical operators.
Note that if we consider the maximally symmetry breaking operators $\mathcal{O}^m_N$, we are able to  derive a closed-form expression that is exact in $m$, at least for some simple cases. We refer to \autoref{App:4pft} for some examples, whereas in the following we consider generic canonical operators $\cO^m_K$.

\subsubsection{HHLL correlators for canonical operators}\label{sec:intcanopm}

 We now analyze the integrated correlators for the canonical operators \(\mathcal{O}^m_K \), which we denote as $\mathcal{G}_{m,K}$ using supersymmetric localization, from which we will deduce the corresponding HHLL correlators.  
More explicitly, we can express \eqref{eq:22mm_example} as: 
\be\label{eq:integ_us}
\mathcal{G}_{m,K}(\tau_2) = \dfrac{\int d^{N-1}a \, \prod_{1 \leq i<j \leq N} a_{ij}^2 \, \partial_\mu^2 Z_{\textrm{1-loop}} \, e^{- \mathrm{Tr} \, a^2} \, \mathcal{O}^m_K \mathcal{O}^m_K }{\int d^{N-1}a \, \prod_{1 \leq i<j \leq N} a_{ij}^2 \, e^{- \mathrm{Tr} \, a^2} \, \mathcal{O}^m_K\mathcal{O}^m_K} \bigg|_{\mu=0}\, .
\ee
Using the decomposition \eqref{eq:z1loop_pert} for $\partial_\mu^2 Z_{\textrm{1-loop}}|_{\mu=0}$, we can immediately express the weak coupling expansion  \eqref{eq:integ_us} in terms of three-point functions: 
\begin{align}   \label{eq:4pt-3pt}\mathcal{G}_{m,K}(\tau_2) = -2 \sum_{L=1}^\infty\sum_{j=0}^{2L}(-1)^{L+j} \left(\frac{1}{2\pi\tau_2}\right)^L \binom{2L}{j}\,(2L{+}1)\zeta(2L{+}1){\llangle \mathcal{O}^m_K(a)  \mathcal{O}^m_K(a) \, t_{2L -j,j}(a)\rrangle  \over  \llangle \mathcal{O}^m_K(a)  \mathcal{O}^m_K(a)\rrangle } \, . 
\end{align}
%
Expressing the integrated correlators in terms of a sum of three-point functions is extremely useful.\footnote{This idea was first introduced in \cite{Brown:2024tru} to compute a different type of integrated heavy-heavy-light-light correlator, where the heavy operators are giant gravitons (i.e., determinant operators).}  Indeed, we can now simply apply here the expressions of three-point functions in the large-charge 't Hooft limit as given in \eqref{eq:3pres}. From the explicit three-point functions, we see that their $L$-loop contribution behaves as $m^L$, which combines nicely with the factor $\tau_2^{-L}$ to form the large-charge 't Hooft coupling $\lambda$ defined in \eqref{eq:thin}. After using the $SU(N)$ trace relations and explicit expressions of three-point functions, from \eqref{eq:4pt-3pt} we obtain 
\be\label{eq:Gweak}
  {\mathcal G}_{m,K}(\lambda) \simeq 4 \sum^{\infty}_{L = 1} (-1)^{L+1} \frac{\Gamma(2L+2)}{\Gamma(L+1)^2} \zeta(2L{+}1) \left[ \sum_{1 \leq i < j \leq K} \left(4 \sin^2 \frac{\pi(i{-}j)}{K} \lambda \right)^L + K(N{-}K)  \lambda^L \right]\, .
\ee
We can already observe that the two coefficients inside the brackets in \eqref{eq:Gweak} precisely correspond to the masses of the fluctuations around the semiclassical configuration in \eqref{eq:aijmass0} and \eqref{eq:aijmass}, with the overall factors reflecting the multiplicities.  We can therefore express \eqref{eq:Gweak} in terms of masses given in \eqref{eq:Ms}, and write it in an extremely simple form,  
\be \label{eq:Gweak2}  
{\mathcal G}_{m,K}(\lambda) \simeq - 2 \sum^{\infty}_{L = 1}  \frac{\Gamma(2L+2)}{\Gamma(L+1)^2} \zeta(2L+1)  \sum_{s} \left(- M_{s}^2  \right)^L  \, . 
\ee
With the results of the integrated HHLL correlators of the canonical operators from supersymmetric localization as given in \eqref{eq:Gweak2}, it is straightforward to determine the color factor $d_{\cH, N;L}$ in \eqref{eq:Tuv_gen}, and therefore determine the full four-point dynamical correlator $\mathcal T_{m,K}(u, v; \lambda)$. 

By comparing with \eqref{eq:gH-Feyn}, we find the color factor for HHLL correlators of canonical operators given as, 
\ie \label{eq:ddd}
d_{\cO_K^m, N;L} =    {K \over 2}\sum_{s=1}^{K-1} \left(4 \sin^2 \left({s \pi \over K} \right)  \right)^L +K (N-K)    \, .
\fe
Plugging the above color factor back into the  general expression for HHLL correlators of generic heavy operators given in \eqref{eq:Tuv_gen}, we then find that the HHLL correlators of the canonical operators take the following form
\ie \label{eq:Tuv_gen3}
    \mathcal T_{m,K}(u, v; \lambda) = K  \sum_{s=1}^{K-1} L(u,v; 4 \lambda \sin^2 {\pi s \over K} ) + 2 K (N-K) L(u,v; \lambda ) \, , 
\fe
where 
\ie
L(u,v;a) = {1 \over 2u} \left[ \left( \sum_{L=0}^{\infty} (-a)^{L} P^{(L)} (u,v) \right)^2 -1 \right] \, . 
\fe
This expression is identical to the one given in \eqref{eq:Tuv_final_clas} obtained from the semiclassical analysis. 
This agreement provides consistent evidence supporting the conjectured classical profile for the scalars in the presence of the heavy canonical operators, as given in \eqref{eq:Zclass2} and \eqref{eq:Zclass}.

\subsection{Extended analysis of integrated correlators}

In this subsection we further investigate the integrated correlators of canonical operators and provide some additional results in that case. In particular, we consider the integrated correlators beyond perturbation theory, and we examine their universal properties in the large-charge 't Hooft limit with $N$ being large as well. We also discuss the dual matrix model description for the integrated correlators by extending the results of \cite{Grassi:2024bwl}.

\subsubsection{Strong coupling analysis}

Starting from \eqref{eq:Gweak2}, the sum over the loops $L$ is convergent with a finite radius, which allows for the resummation and strong-coupling expansion by using the identity (see e.g. \cite{Grassi:2019txd}): 
\be\label{eq:idenitoy} 
\ba
& - \sum_{L \ge 1}  \frac{\Gamma\left(2L+2\right)}{\Gamma\left(L+1\right)^2} \zeta(2L+1) (-a)^L =  
\int_0^\infty dw \frac{w}{\sinh^2w} \left( 1-J_0  \left( 4 w \sqrt{a} \right) \right) \cr
=&\, 1 + \gamma_E + {\log a \over 2}  + 2 \sum_{n\ge1} \left(4\pi n \sqrt{a} K_1 \left(4 \pi n \sqrt{a}\right) - K_0  \left(4 \pi n \sqrt{a}\right)\right) \, . 
\ea
\ee
This leads to 
\be\label{eq:4ptstrong}
\ba
 {\mathcal G}_{m,K}(\lambda) & \simeq  \,  2K(2N-K-1)( 1 + \gamma_E )+ 2 K\left(\log K+\left(N-\frac{K+1}{2}\right)\log \lambda\right)
 \cr 
 & + 4 \sum_s  \sum_{n\ge1} \left(4\pi n M_s K_1 \left(4 \pi n M_s \right) - K_0  \left(4 \pi n M_s\right)\right) \, , 
\ea
\ee
where the explicit expression for $M_s$ is given in \eqref{eq:Ms}, and we used equation \eqref{eq:summssq}. 
The terms in the second line in the above expression are fully non-perturbative at large \(\lambda \), as the modified Bessel functions \(K_1\) and \(K_0 \) exhibit the asymptotic behavior:
\be  \label{def:as}
  K_{\nu}(4\pi n M_s) \sim \re^{-4\pi n M_s} \, .
\ee
Such exponentially small effects correspond to the worldline instanton action of a massive  particle, with mass $M_s$, on  $S^3 \times \mathbb{R}$, which winds around the equator of  $S^3$.
From this perspective, the index  $n$  in the summation represents the winding number \cite{Grassi:2019txd,Hellerman:2021duh}.

\subsubsection{Large $N$ and large charge}\label{sec:largeN}

We now can study the case where $N$ is also taken to be large, with a precise ranking between the two limits: our analysis is valid when the dimension of the heavy operator is much larger than $N^2$. 
In this subsection, we explore the large-$N$ limit of the integrated HHLL correlators $\mathcal{G}_{m,K}$.
Let us consider the coupling $\lambda$ within the radius of convergence, then the weak coupling expansion is absolutely convergent and we can therefore exchange the limit with the sum over $L$ in \eqref{eq:Gweak}. We consider two cases separately. 

First, we consider the limit, 
\be \label{eq:largeN1}  N\to \infty,\quad\text{with}\quad K \quad   \text{fixed}\, . \ee
The first term in \eqref{eq:Gweak} in this limit drops out, and the analysis proceeds in a similar way. In particular, the structure of the correlators does not change drastically and we have 
\be  \mathcal{G}_{m,K} (\lambda)\Big|_{\text{\eqref{eq:largeN1}}}\simeq 4 NK \sum_{L \geq 1} (-1)^{L+1} \frac{\Gamma(2L+2)}{\Gamma(L+1)^2} \zeta(2L+1)  \lambda^L ~.\ee
The resummation in $\lambda$ can be obtained directly using \eqref{eq:idenitoy}.

In the second case we take 
\be \label{eq:largeN2} N ,K \to \infty,\quad\text{with} \quad  \kappa={K\over N},\quad \text{fixed}\,.\ee
In this case the structure is different. 
Let us first recall that
\be\ba \label{eq:largeK} {1\over K} \sum_{s=1}^{K-1}  \left(4\sin^2\left({\pi s \over K}\right)\right)^L&=\int_{0}^1\rd x (4\sin^2\left(\pi x\right))^L+\mathcal{O}\left({1\over K}\right)
=\frac{2^{2L} \Gamma \left(L+\frac{1}{2}\right)}{\sqrt{\pi } \Gamma (L+1)}+\mathcal{O}\left({1\over K}\right)~.\ea\ee
Using this identity we have, at leading order in the 't Hooft expansion  and at large $N$,
\be  \label{eq:largeweak}\ba \mathcal{G}_{m,K}(\lambda)\Big|_{\text{\eqref{eq:largeN2}}}\simeq & \, - 4 \kappa  N^2 \sum _{L=1}^{\infty } \frac{ (-\lambda) ^L \zeta (2 L{+}1) \Gamma \left(L{+}\frac{3}{2}\right) \left[ \sqrt{\pi }\, 2^{2L+1} (1 {-} \kappa)  \Gamma (L{+}1) + 2^{4L}\, \kappa  \, \Gamma \left(L{+}\frac{1}{2}\right)\right]}{\pi  \Gamma (L+1)^2}\\
=& \, 4 \kappa  N^2 \! \int_0 ^{\infty} \!  \rd w \frac{w}{\sinh ^2(w)}   \left[  (1 - \kappa )  \left(1-J_0 \left(4 w \sqrt{\lambda } \right)\right) +  {\kappa \over 2}    \left(1- J_0 \left(4 w \sqrt{\lambda } \right){}^2\right) \right]~. 
\ea\ee
The same result can be obtained if we first resum the perturbative series \eqref{eq:Gweak2} using \eqref{eq:idenitoy}, and then take the large $N$ and large $K$ limit. From this result, it is immediately clear that, in this limit, the color factor no longer takes the form $c^L$ in general; instead, we have an entirely new structure as compared to finite $K$.
Interestingly, for $\kappa = 1$, up to a redefinition of $\lambda$, this expression precisely matches \cite[eq.~(6.80)]{Grassi:2024bwl}. Therefore it appears that the maximal trace component inside the canonical operators $\mathcal{O}^m_K $ dominates at large $N=K$ in the scaling \eqref{eq:largeN2}. Differences occur if $\kappa\neq 1$. 

As before, the analytic continuation at large $\lambda$ can be performed using the Mellin-Barnes transform, yielding: 
\be\label{eq:largestrong}\ba  \mathcal{G}_{m,K}(\lambda)\Big|_{\text{\eqref{eq:largeN2}}}\simeq &\,  (2-\kappa) \kappa  N^2 \left(\log \left(\lambda\right)+2 \gamma_E +2\right) \, + \, 16 \kappa ^2  N^2  \sum_{k\geq 1} {a_k\over \lambda^{k-1/2} } \\
& +  8 N^2\kappa(1-\kappa) \sum_{n\ge1} \left(4\pi n \sqrt{\lambda} K_1 \left(4 \pi n \sqrt{\lambda} \right) - K_0  \left(4 \pi n \sqrt{\lambda} \right)\right) \, , \ea\ee
with
\be a_k=  \frac{\zeta (2 k-1) \Gamma \left(k-\frac{1}{2}\right)^3}{\sqrt{\pi} (16 \pi^2 )^{k} \Gamma (k-1)}~.\ee
This is very different from what we have at finite $N$ or in the limit \eqref{eq:largeN1}, where there are no $1/\lambda$ perturbative terms. 
The emergence of a new perturbative series  in $1/\lambda$ was explained in \cite{Grassi:2024bwl}  by identifying a specific instanton action that vanishes in the relevant limit. Likewise, here for the instanton actions in \eqref{def:as} we have $\lim_{K\to \infty} M_s= \lim_{K\to \infty} 2 \sin\left(\pi s\over K\right)=0$ for $s$ finite, indicating that a large number of BPS particles become massless in this scaling limit.

In addition we have 
\be a_k\sim \Gamma(2k-1) \pi^{-2k}\left(2^{4-10 k}+2^{3-8 k}+2^{2-6 k}+2^{2-6 k} 3^{1-2 k}+\mathcal{O}(1/k)\right) \, ,\ee
and we can deduce that \eqref{eq:largestrong} is an asymptotic expansion with instanton actions given by integers multiples of $8 \pi$,  
which is independent of $N$ and  agrees with the SU(3) example \cite[eq.(6.84)]{Grassi:2024bwl} as expected from the discussion above.
Note also that the \( 1/\lambda \)  expansion in \eqref{eq:largestrong} arises entirely from the last term in \eqref{eq:largeweak}. The term proportional to \( 1 - \kappa \) does not contribute to the \( 1/\lambda \) expansion.
By performing a median Borel summation (see e.g. \cite{Dorigoni:2014hea, Aniceto:2018bis} for a review) parallel to \cite{Grassi:2024bwl}, we find that \eqref{eq:largestrong} reproduces the full expression given in the second line of \eqref{eq:largeweak}.

\subsubsection{Large charge matrix model}\label{sec:matrixmodelrepr}
In \cite{Grassi:2024bwl} it was found that integrated correlators for rank 2 canonical operators are described by a Jacobi matrix model whose eigenvalue density is
\be \sigma(x)={1\over \pi}{1\over \sqrt{x(1-x)}}~.\ee
The rank of the matrices in such Jacobi model is related to the charge $mK$.  
This makes the matrix model representation of \cite{Grassi:2024bwl}\footnote{Such matrix models are completely different from the one discussed in \autoref{sec:localization}. and have also been denoted as ``dual matrix models'' \cite{Grassi:2019txd,Beccaria:2020azj,Caetano:2023zwe} with respect to the Pestun's matrix model.} particularly well-suited for describing the large-charge behavior of the correlators.
In the large charge 't Hooft limit then the integrated correlator is then given by
\be
 {\mathcal G}_{m,3}^{SU(3)}(\lambda)=\int_0 ^1 \left(\partial^2_{\mu}Z_{\rm 1-loop} \right)\Big|_{\mu=0,~ {\rm Tr}a^2= 6 \lambda, ~ {\rm Tr}a^3=6 \lambda ^{3/2} (2 x-1) }\, \sigma(x)\, {\rd x} \,.
\ee
Similarly, it turns out that we can express \eqref{eq:Gweak2} as the integral over a Jacobi density 
\be   {\mathcal G}_{m,K}(\lambda)=\int_0 ^1 \left(\partial^2_{\mu}Z_{\rm 1-loop}\Big|_{\mu=0}\right)\Big|_{\rm locus} \sigma(x)\, {\rd x} \, , \ee
where the locus is specified by the value of the first $K$ independent traces, which are: 
\be\label{eq:locus}\ba
\Tr a^m=2 K \lambda^{\frac{m}{2}}  \left( \binom{m-1}{m/2} \epsilon_m + \left(2x-1 \right) \delta_{m,K} \right)\, , \quad m=2,\dots, K~,
\ea\ee
where $\epsilon_m=1$ if $m$ is even, and $0$ if $m$ is odd. To make a direct contact with \eqref{eq:3pres},  note that \eqref{eq:locus} can also be written as 
\be \ba \Tr a^m= \lambda^{\frac{m}{2}} \sum_{s=0}^{K-1} \left(\re^{-\frac{2 \ri \pi  s}{K}} \left(\sqrt{x-1}-\sqrt{x}\right)^{-2/K}+\re^{\frac{2 \ri \pi  s}{K}} \left(\sqrt{x-1}-\sqrt{x}\right)^{2/K}\right)^m~.\\
\ea\ee
For the rank 2 case, the Jacobi matrix model  exactly describes the mixing structure of $t_3$ and $t_2$ on the sphere, and hence it contains the all orders $1/m$ corrections. On the other hand, for $N>3$ the mixing structure should be more complicated, therefore it is not clear if the full $1/m$ expansion is captured by a Jacobi model.

\section{Conclusion and outlook}\label{sec:conc}

In this paper we explored the dynamical properties of the large-charge sector of $\mathcal{N}=4$ SYM. 
Specifically, we examined the large-charge limit of a class of correlation functions involving a particular type of heavy operator, referred to as a ``canonical operators". These insertions map the theory to the Coulomb phase and breaks the $SU(N)$ gauge group. We demonstrated that correlators involving two canonical operators and light operators are fully determined in the large-charge limit by assigning specific classical profiles to the scalars. This was explicitly demonstrated for three- and four-point correlators, where the large charge expressions exhibit remarkably elegant structures. In particular, the HHLL correlators — comprising two canonical operators and two light operators — can be determined exactly, in terms of an infinite sum of ladder Feynman integrals which can be further resummed, yielding an exact result valid at finite values of the coupling constant.

Among the canonical operators, we identified a particularly special one: the maximally symmetry breaking  operator with $K=N$. Notably, the perturbative $1/m$ expansion of their two-point function can be resummed, and its behavior closely resembles that of rank-1 theories at all orders in the large-charge expansion. This suggests that the EFT description of rank-1 correlators proposed in \cite{Hellerman:2017sur,Hellerman:2018xpi} should still apply in this specific case. Deriving this result from an EFT perspective directly and extending it to higher-point functions represents an immediate direction for future work.

Our findings open up numerous avenues for future research. 
Below, we outline a few of them.
\begin{itemize}
\item[-]  Within the context of $\mathcal{N}=4$ SYM, a natural direction is to explore the large-charge limit of operators with reduced supersymmetry. Heavy operators preserving a lower number of supercharges are of great interest due to their connection to black hole physics. Although they are not suited for localization analysis, a semiclassical approach may still be viable.

\item[-] An interesting open direction is the construction of canonical operators in less supersymmetric theories, initiated in \cite{Grassi:2024bwl} for rank-2 $\mathcal{N}=2$ theories. In particular it would be interesting to see how the intricate structure of Coulomb branches of $\mathcal{N}=2$ translates into correlation functions of canonical operators.

\item[-]
From a more technical point of view, one could explore different ways of taking the charge to be large. In this work we considered operators of the form \(\mathcal{O}_{(m_1, m_2, m_3,...)}\), where only one \(m_k \to \infty\) while the others \(m_{i\neq k}\) remain fixed. Generalizing this to cases where multiple quantum numbers become large simultaneously, or considering correlators containing two canonical operators with different values of $K$, would be an interesting direction to explore.

\item[-] Another promising direction is to investigate the emergence of dual large charge matrix models associated to canonical operators. In \cite{Grassi:2024bwl} it has been explicitly shown that canonical correlators in rank 2 theories are described by a Jacobi matrix model, and the considerations in \autoref{sec:matrixmodelrepr} suggest that a similar structure should be present at any rank. This would provide a convenient framework to extend the analysis of the large-charge 't Hooft limit beyond the leading order, considering the large charge limit at fixed gauge coupling, and considering more general large charge limit as discussed above.

\item[-] A well-known property of $\mathcal{N}=4$ SYM is its conjectured S-duality~\cite{Montonen:1977sn, Goddard:1976qe}. Studying this non-perturbative duality necessitates an analysis of correlators at finite Yang-Mills coupling. The modular properties of integrated HHLL correlators, where large-charge operators are maximal-trace operators or their analogous generalizations, have been investigated in~\cite{Brown:2023why, Paul:2023rka} by considering the large-charge expansion with Yang-Mills coupling $\tau$ being fixed. These studies leveraged recursive Laplace-difference equations derived from the supersymmetric localization matrix model~\cite{Brown:2023cpz, Paul:2022piq}. It would be highly valuable to explore the $SL(2, \mathbb{Z})$ modular properties of the correlators involving canonical operators in the large-charge limit at fixed Yang-Mills coupling (see  Appendix \ref{App:4pft} for further comments). 
Understanding S-duality will also aid in analyzing subleading orders in the large-charge 't Hooft limit, since modular symmetry interrelates different orders in the large-charge expansion.

\item[-] 
Another interesting direction of investigation is to understand our results from the point of view of the dual large $N$ geometry \cite{Lin:2004nb}.  Recent developments \cite{Aprile:2024lwy,Turton:2024afd,Aprile:2025hlt} have shown that certain four-point correlators of superconformal primary operators in $\mathcal{N}=4$ SYM can be derived by viewing them as two-point functions in a non-trivial geometry of the holographic theory (see also \cite{Buchbinder:2010ek} for earlier work in computing holographic HHLL correlators).  It is of great interest to explore whether the large-$N$ behavior of the canonical operators in the large-charge limit has a nice geometric interpretation in the dual holographic theory. As we have seen in the main text in \autoref{sec:largeN}, our results exhibit universal structures in the large $N$ limit, which strongly suggests such an interpretation is possible.  
In this context, it would be very interesting to understand the  gravity interpretation of the OPE data \eqref{eq:heavyOPE} and establish a direct connection with recent developments in the light-cone bootstrap \cite{Kulaxizi:2018dxo, Karlsson:2019qfi,Karlsson:2021duj}.

\item[-]  Within this holographic context, it would also be important to make contact with the gauge symmetry breaking patterns induced by special brane constructions, along the lines of \cite{Skenderis:2006di,Ivanovskiy:2024vel}, representing important examples of non-conformal holography. Analogously, our results on the Coulomb branch - both from semiclassics and from localization - could represent important constraints for the non-conformal bootstrap \cite{Karananas:2017zrg,Cuomo:2024fuy}.

\end{itemize}

\section*{Acknowledgements}
We thank Joao Caetano, Gabriel Cuomo, Zohar Komargodski, Shota Komatsu, Mukund Rangamani, Rodolfo Russo, Raffaele Savelli, Luigi Tizzano for helpful discussions. We would like to thank Shota Komatsu, Rodolfo Russo, and Luigi Tizzano for useful comments on the draft. 
AB is supported by a Royal Society funding RF$\backslash$ERE$\backslash$210067. CW is supported by a Royal Society University Research Fellowship,  URF$\backslash$R$\backslash$221015. FG and CW are supported by a STFC Consolidated Grant,  ST$\backslash$T000686$\backslash$1 ``Amplitudes, strings \& duality".  The work of AG and CI  is partially supported by the Swiss National Science Foundation Grant No. 185723 and the NCCR SwissMAP. No new data were generated or analyzed during this study.

\appendix

\section{Gaussian matrix models techniques}\label{app:Gauss_MM}
In this appendix we detail the  techniques that allow to perform computations in the Gaussian matrix model defined in \eqref{eq:Z_matrixModelGaussian} at fixed rank, following \cite{Billo:2017glv,Beccaria:2020hgy,Galvagno:2020cgq} where this set of recursive techniques were developed.
We rewrite the Gaussian partition function as:
\begin{equation}\label{eq:GaussZ_matrix}
    \cZ=\int da~{\text e}^{-\tr a^2}~,~~~~~ a=\sum_{b=1}^{N^2-1} a^b\,T_b\,,~~~ da = \prod_{b=1}^{N^2-1} \frac{da_b}{\sqrt{2\pi}}\,,
\end{equation}
where $a$ is a $N\times N$ Hermitian matrix which takes value in the $\mathfrak{su}(N)$ gauge algebra and the integration measure is normalized such that $\cZ=1$.
The $\mathfrak{su}(N)$ generators are normalized as: 
\begin{align}\label{eq:sun_normalis}
    \tr\,T_{b}\,T_{c} =\frac{1}{2}\,\delta_{bc}\,,~~~~~\tr T_b = 0\,.
\end{align}
Any gauge invariant observable inserted in this matrix model as in \eqref{eq:f_Gaussian} reads:
\begin{align} \label{eq:gauss2}
    \llangle f(a)\rrangle = \int da~{\text e}^{-\tr a^2} f(a)\,.
\end{align}
The generic function $f(a)$ is expected to be written in terms of traces of powers of the matrix $a$, for which we conveniently introduce the notation:
\begin{equation}\label{eq:t_def}
     u_{\bs p} = \llangle\tr a^{p_1} \tr a^{p_2} \dots \tr a^{p_m}\rrangle\,,
\end{equation}
where $m$ is the length of the vector $\bs p$. Expectation values of multitrace insertions \eqref{eq:t_def} can be evaluated at fixed $N$ through the basic Wick contraction $\vev{a_b\, a_c}_0 = \delta_{bc}$, and employing the following $\mathfrak{su}(N)$ matrix fission/fusion identities for arbitrary $N\times N$ matrices $B_1$ and $B_2$:
\begin{equation}\label{eq:fiss_fus}
    \begin{split}
    \tr T^b B_1 T^b B_2 &= \frac{1}{2} \tr B_1 \tr B_2 -\frac{1}{2N} \tr B_1 B_2\,, \\
    \tr T^b B_1 \tr T^b B_2 &= \frac{1}{2} \tr B_1 B_2 -\frac{1}{2N} \tr B_1 \tr B_2\,.
\end{split}
\end{equation}
This set of rules can be implemented using a recursive routine. Starting from the $\mathfrak{su}(N)$ initial conditions ensuring \eqref{eq:sun_normalis}:
\begin{align}
	\label{eq:init_cond}
		u_p = 0 \, ,  \quad \text{for \,  $p$ \,  odd}\, , \qquad \text{and} \qquad u_0=N\,,
\end{align}
and using \eqref{eq:fiss_fus}
one can evaluate the general combination $u_{\bs p}$ in terms of a rational function in $N$ by implementing the following recursive formulas:
\begin{align}
\begin{split}\label{eq:recursion_form}
u_{p_{1},p_{2},\dots,p_{m}} =   &\, \frac{1}{2} \sum_{j=0}^{p_{1}-2}  \left( u_{j,p_1-j-2,p_2,\dots,p_m} \right)
		-\frac{p_1-1}{2N}\,   u_{p_{1}-2,p_{2},\dots,p_{m}} \\ 
		+& \sum_{k=2}^{m}\frac{p_{k}}{2} \,\Big(  u_{p_1+p_k-2,p_2,\dots,\slashed{p_k},\dots,p_m} -\frac{1}{N} \,u_{p_1-1,p_2,\dots,p_k-1,\dots,p_m} \Big)\,,
\end{split}
\end{align}
where the notation ${p_1, \ldots ,\slashed{p_k}, \ldots ,p_m}$ represents the sequence of indices without the $k$-th one. Some explicit results for generic $N$ follow:
\begin{equation}
\begin{split}
&u_{2}=\frac{N^2-1}{2}\,, ~~~~~
u_{4}=\frac{(N^2-1)(2N^2-3)}{4N} \,, \\
&u_{2,2}=\frac{N^4-1}{4}\,, ~~~~~
u_{6}=\frac{5(N^2-1)(N^4-3N^2+3)}{8N^2} \, , \\
&u_{4,2}=\frac{(N^2-1)(N^2+3)(2N^2-3)}{8N}\,,~~~~~
u_{3,3}=\frac{3(N^2-1)(N^2-4)}{8N}\,.
\end{split}
\end{equation}
This procedure can be implemented in {\tt Mathematica} and greatly simplifies the computations in the matrix model for a generic $N$.

\section{Examples of three-point functions}\label{App:threept}

In this appendix, we provide a few examples of explicit computation of three-point functions that match precisely the predictions of semiclassical analysis, namely \eqref{eq:threeptodd} (for odd $K$) and \eqref{eq:threepteven} (for even $K$).

We first consider an example of maximally symmetry breaking operators, with $N=5$, $K=5$, and $t_{\bold{p}}$ with $\bold{p}=(4)$. As discussed in \autoref{sec:3pnt}, the three-point functions for $K=N$ can be found for all charges.  By calculating up to $m=15$, we find  
\begin{align} \label{eq:NK5}
\frac{\llangle \mathcal{O}_5^m \mathcal{O}_5^m t_{4}\rrangle }{\llangle \mathcal{O}_5^m \mathcal{O}_5^m \rrangle} 
&= \, \frac{30}{2^{m+2} 5^{3 m+5}+ \left(1+(-1)^m\right) \, 2^{m+2} 5^{m/2+1} -  (x^+_{5})^{m+2}-(x^-_{5})^{m+2}} \\
   &\bigg[ 2^m 5^{3 m+4} \left(5 m^2+ 27 m-2\right) -{(x^+_{5})^m+(x^-_{5})^m \over 2}  \left( 325 m^2+2219
   m - 1290 \right) \cr
   &- \sqrt{5} {(x^+_{5})^m - (x^-_{5})^m \over 2}
   \left( 145 m^2 + 991 m - 578 \right) \cr
   &+2^m 5^{m/2} \left( \left(1+(-1)^m\right) (8 m-20)-  { 1-(-1)^m \over \sqrt{5}} 
 (25m^2+95m+90)  \right)\bigg]\, ,    \nonumber
\end{align} 
where $x_5^{+}$ and $x_5^{-}$ are solutions to 
\ie \label{eq:solx5}
 x^2- 50 x +20 = 0 \, ,  \quad {\rm with} \quad x_5^+ = 25 +11 \sqrt{5}\, , \quad x_5^- = 25 -11 \sqrt{5} \, . 
\fe 
In the large-$m$ expansion, the expression given in \eqref{eq:NK5} reduces to $15m^2/2$, which agrees with the prediction from \eqref{eq:threeptodd}. 

\begin{figure}[t!]
    \centering
    \includegraphics[width=0.6\linewidth]{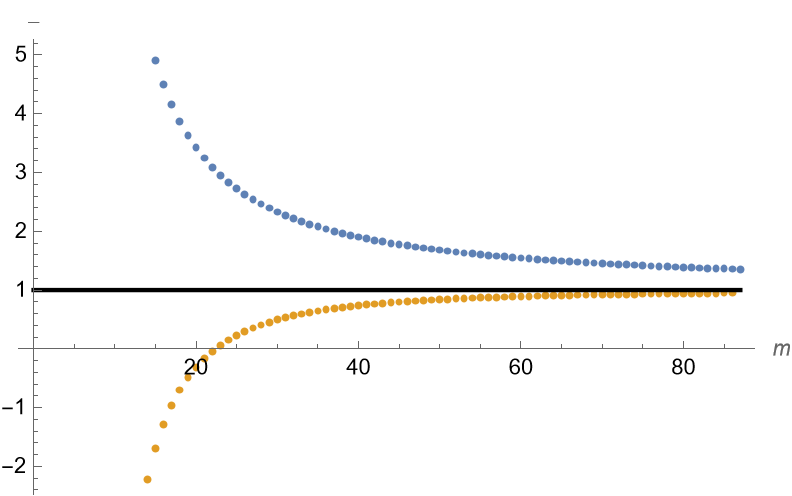}
    \caption{The numerical data for $\text{\eqref{eq:b33}}/\big(\frac{729}{16}m^7\big)$ is in blue, while its Richardson extrapolation is in orange.}
    \label{threepoint1}
\end{figure}

We can consider more general canonical operators. For example, for $N=4$, $K=3$, and $\bold{p}=(4,4,3,3)$. In this case, we calculate up to $m=70$ and apply the Pade method to find
\ie\label{eq:b33}
\frac{\llangle \mathcal{O}_3^m \mathcal{O}_3^m t_{4^2,3^2}\rrangle }{\llangle \mathcal{O}_3^m\mathcal{O}_3^m \rrangle } = 45.5625\, m^7 + O(m^6)\, , 
\fe
which, after applying the $SU(3)$ trace relation $t_4= 1/2\, t_{2,2}$, agrees with the prediction from \eqref{eq:threepteven} that is  $729m^7/16$. The explicit numerical data is demonstrated in Fig. \ref{threepoint1}. 

\section{Examples of four-point functions}\label{App:4pft}

From \eqref{eq:Gweak2}, we predict, in the large-charge limit, the integrated four-point functions to be given by 
\be \label{eq:Gweak2app}  
{\mathcal G}_{m,K}(\lambda) \simeq 2 \sum^{\infty}_{L = 1} (-1)^{L+1} \frac{\Gamma(2L+2)}{\Gamma(L+1)^2} \zeta(2L+1)  \left[K \sum_{s=1}^{K-1} \left(4\lambda \sin^2 \frac{\pi s}{K}  \right)^L + 2K(N-K) \left(\lambda \right)^L \right] \, . 
\ee
As a first test, we can consider $SU(3)$. We consider the most general operator 
\begin{align}\label{eq:fullSU34point}
    &\mathcal{O}_{3^{m_3} 2^{m_2}}  = t_3^{m_3}t_2^{m_2} + \text{GS}\, ~\, {\rm with} \quad \{ t_3^{n_3} t_2^{n_2} \}_{\Delta=3n_3+2n_2,{n_3\le m_3}} \cup \{ t_3^{n_3} t_2^{n_2} \}_{\Delta \le 3m_3+2m_2} \, ,
\end{align}
(see \autoref{sec:canonical} for more examples of the Gram-Schmidt procedure). For these operators, we find, checking up to $m_3=100$ and $L=10$,  the full integrated four-point function
\begin{align}
\mathcal{G}_{3^{m_3} 2^{m_2}}(\tau_2) =& \, 2 \sum_{L=1}^\infty (-1)^{L} \frac{\Gamma(2L+2)}{\Gamma(L+1)} \left(\frac{1}{4 \pi \tau_2} \right)^L \zeta(2L+1)\bigg((2L+1)(L^2+L+6) \\
&+6 \, _3F_2\left( -L,L+1,-m_2; 1, \frac{15}{2}+3m_3;1 \right) \frac{(-L)_{(3m_3+3)}-(L+1)_{(3m_3+3)}}{\Gamma(3m_3+4)}\bigg) \, . \nonumber
\end{align}
It is interesting to note that the summand is invariant under $L \rightarrow -L-1$, which is related to the fact that the correlator should be $SL(2,\mathbb{Z})$ invariant. Indeed, following the same assumption of \cite{Dorigoni:2021bvj,Dorigoni:2021guq, Paul:2022piq, Paul:2023rka, Brown:2023why, Brown:2023cpz}, one may promote the above expression to be manifestly $SL(2,\mathbb{Z})$ invariant by writing it in terms of non-holomorphic Eisenstein series. We can also study its large-charge properties while maintaining the manifest $SL(2, \mathbb{Z})$
invariant by keeping $\tau$ fixed, as in \cite{Brown:2023why, Paul:2023rka}. 

To compare to \eqref{eq:Gweak2app},  we take $m_3$ large with fixed $\lambda$  and $m_2$, and find
\ie \label{eq:su3largem}
\mathcal{G}_{m,3} (\lambda) \simeq 4 \sum_{L=1}^\infty &(-1)^{L+1} \frac{\Gamma(2L+2)}{\Gamma(L+1)^2} 3 \left(3 \lambda \right)^L \, .
\fe
We see that \eqref{eq:su3largem} agrees with the prediction of \eqref{eq:Gweak2app}. Note that, as explained in \autoref{sec:canonical}, the $m_2$ dependence drops out in the leading large $m_3$ limit.

Let us consider another case with $N=5$, $K=5$. As discussed in \autoref{sec:3pnt}, for such maximally symmetry breaking operator, we can determine the integrated correlator for all charges (order by order in perturbation). By calculating up to $m=20$,  we find the result for the first three loops: 
 \allowdisplaybreaks{
\begin{align}
&\mathcal{G}_{m,5}(\tau_2) \bigg|_{L=1} =\frac{150 \zeta(5)}{\pi \tau_2}m \, ,\\
&\mathcal{G}_{m,5}(\tau_2) \bigg|_{L=2} =-\frac{4500 \zeta (5)}{\pi^2  \tau_2^2\, \left( 2^{m+2} 5^{3 m+5}+2^{m+2} 5^{m/2+1} \left(1+(-1)^m\right)- (x_5^+)^{m+2}- (x_5^-)^{m+2} \right) } \cr
   &\Bigg[ 2^{m}\, 5^{3 m+4} \left(\frac{5m^2}{2}+\frac{77m}{6}-\frac{1}{3} \right) - \big(\left(x_5^+\right)^m+\left(x_5^-\right)^m \big) \frac{\left(1555m^2+8369m-1290\right)}{12}  \cr
   &+2^m 5^{m \over 2} \bigg(\left(1+(-1)^m\right) \frac{\left(5m^2+29m-10\right)}{3} -  \sqrt{5}\left(1-(-1)^m\right) \left(\frac{5m^2}{6}+\frac{19m}{6}+3\right)  \bigg)
    \cr
   &- \sqrt{5} \big( \left(x_5^+\right)^m-\left(x_5^-\right)^m \big)
   \left(\frac{695 m^2}{12}+\frac{1247 m}{4}-\frac{289}{6}\right) \Bigg]  \, , \\
   &\mathcal{G}_{m,5}(\tau_2) \bigg|_{L=3} =\frac{220500 \zeta (7)}{\pi^3  \tau_2^3\, \left( 2^{m+2} 5^{3 m+5}+2^{m+2} 5^{m/2+1} \left(1+(-1)^m\right)- (x_5^+)^{m+2}- (x_5^-)^{m+2} \right) } \cr
   &\Bigg[ 2^{m}\, 5^{3 m+4}\frac{\left(50 m^3+405 m^2+1087 m-84 \right)}{252} - \big(\left(x_5^+\right)^m+\left(x_5^-\right)^m \big) \frac{ 50m^3+1825+7623-5840 }{56}  \cr
   &+2^m 5^{m \over 2} \bigg(\big(1+(-1)^m \big) \left(\frac{25 m^3}{72}{+}\frac{125 m^2}{56}{+}\frac{1279 m}{252}{-}\frac{5}{42}\right) {-} 5 \sqrt{5} \big( 1-(-1)^m \big)
    \frac{5m^3{+}29m^2{+}56m{+}36 }{56} \bigg)
    \cr
   &- 5\sqrt{5} \big(\left(x_5^+\right)^m-\left(x_5^-\right)^m \big) 
   \left(\frac{5 m^3}{63}+\frac{163 m^2}{56}+\frac{6131 m}{504}-\frac{28}{3}\right) \Bigg]  \, ,
\end{align}}
\hspace{-0.325cm} where $x_5^{\pm}$ are given in \eqref{eq:solx5}. The expressions show a clear structure, which allow for extensions to higher orders.  In the large-$m$ limit and expressing the results  in terms of $\lambda$, we have
\ie
\mathcal{G}_{m,5}(\lambda) =600 \lambda \zeta(3) -7200 \lambda^2 \zeta(5)+ 140000 \lambda^3 \zeta(7) + \mathcal{O}(\lambda^4,1/m)
\fe
which agrees with the prediction from \eqref{eq:Gweak2app}.

\begin{figure}[t!]
    \centering
    \includegraphics[width=0.6\linewidth]{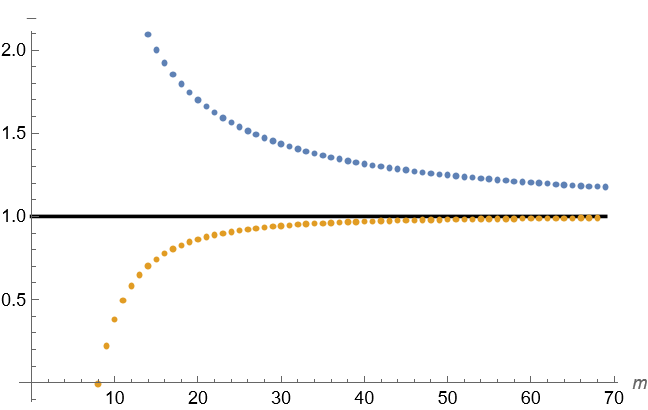}
    \caption{The numerical data for $\mathcal{G}_{m,3}(\tau) \big|_{L=4}\big/\big(\!-619920 \lambda^4  \zeta(9)\big)$ is in blue, and its Richardson extrapolation is in orange.}
    \label{fourpoint2}
\end{figure}

Finally, we consider the case of non-maximally symmetry breaking operators, for example $N=4$, $K=3$. Calculating up to $m=69$ and using the Pade method, we find
\ie
\mathcal{G}_{m,3}(\lambda) &= 288.0 \lambda \zeta(3) - \left( 3600.0 \lambda^2 \zeta(5) + \mathcal{O}(1/m)\right)+\left( 47040.0 \lambda^3 \zeta(7) + \mathcal{O}(1/m)\right) \cr
&-\left(619917.3 \lambda^4  \zeta(9) + \mathcal{O}(1/m)\right) + \mathcal{\lambda^5} \, ,
\fe
which once again agrees with the prediction from \eqref{eq:Gweak2app} of
\ie
\mathcal{G}_{m,3}(\lambda) = 288 \lambda \zeta(3) -  3600 \lambda^2 \zeta(5)+ 47040 \lambda^3 \zeta(7)
-619920\lambda^4  \zeta(9) + \mathcal{O}(\lambda^5) \, .
\fe
In figure \ref{fourpoint2}, we show the numerical data for  the order $\lambda^4$ term, and the convergence of the numerical data that is in agreement with the prediction from the semiclassical analysis.

\bibliographystyle{JHEP}
\bibliography{biblio}

\end{document}